\documentclass[11pt,a4paper]{article}

\usepackage[utf8]{inputenc}

\usepackage{fullpage}
\usepackage{hyperref}

\usepackage{cite}
\usepackage{graphicx}

\usepackage{amsmath}
\usepackage{amsfonts}

\usepackage{subfigure}
\usepackage{xcolor}

\renewcommand{\vec}[1]{\mathbf{#1}}
\newcommand{\myfigurename}{Figure}

\usepackage{enumerate}

\newtheorem{theorem}{Theorem}
\newtheorem{defn}{Definition}
\newenvironment{definition}{\begin{defn}\rm }{\hfill \hspace*{1pt} \hfill $\lrcorner$ \end{defn}}
\newtheorem{assumption}{Assumption}
\newtheorem{exa}{Example}

\newtheorem{remark}{Remark}

\newcommand{\eg}{e.g.}
\newcommand{\ie}{i.e.}

\newcommand{\reveqdef}{=:}

\newcommand{\tr}[1]{#1^T}

\newcommand{\vecF}{\vec{F}}
\newcommand{\vecG}{\vec{G}}
\newcommand{\vecH}{H}
\newcommand{\vecx}{\vec{x}}

\newcommand{\vecp}{\vec{p}}
\newcommand{\vecq}{\vec{q}}
\newcommand{\vece}{\vec{e}}
\newcommand{\vecPhi}{\vec{\Phi}}

\newcommand{\vech}{\vec{h}}

\newcommand{\F}{F}
\newcommand{\firemap}{h}
\newcommand{\Firemap}{\vec{H}}
\newcommand{\couplfun}{\Gamma}

\newcommand*\samethanks[1][\value{footnote}]{\footnotemark[#1]}

\begin{document}

	\title{Kick synchronization versus diffusive synchronization}

	\author{Alexandre Mauroy%
		\thanks{A. Mauroy is with the Department of Mechanical Engineering, University of California Santa Barbara, Santa Barbara, CA 93106, USA.  alex.mauroy@engr.ucsb.edu}%
	, Pierre Sacr\'{e}%
	\thanks{P. Sacr\'{e} and R. Sepulchre are with the Department of Electrical Engineering and Computer Science (Montefiore Institute, B28), University of Li\`ege, 4000 Li\`ege, Belgium. pierre.sacre@ulg.ac.be, r.sepulchre@ulg.ac.be}%
	, and Rodolphe Sepulchre\samethanks[2]%
	}

\date{}
	
\maketitle

\begin{abstract}	
The paper provides an introductory discussion about two fundamental models of oscillator synchronization: the (continuous-time) diffusive model, that dominates the mathematical literature on synchronization, and the (hybrid) kick model, that accounts for most popular examples of synchronization, but for which only few theoretical results exist. The paper stresses fundamental differences between the two models, such as the different contraction measures underlying the analysis, as well as important analogies that can be drawn in the limit of weak coupling.
\end{abstract}

%%%%%%%%%%%%%%%%%%%%%%%%%%%%%%%%%%%%%%%%%%%%%%%%%%%%%%%%%%%%%%%%%%%%%%%%%%%%%%%%
%%%%%%% INTRO %%%%%%%

\section{Introduction} \label{sec:intro}
 
% Synchronization is a pervasive concept in science and engineering. Currently, it is perhaps the most widely studied dynamical concept across systems biology~\cite{Goldbeter:1996uo}, chemistry~\cite{Kuramoto:1984wo}, physics~\cite{Huygens:1673vi,Pikovsky:2001et}, neuroscience~\cite{Hoppensteadt:1997tp,Izhikevich:2007vr}, astronomy~\cite{Blekhman:1988wn}, biology~\cite{Winfree:1980ue,Glass:1988ub}, and engineering~\cite{Strogatz:2003tm,Nijmeijer:2003uy}. Because synchronization involves interconnection at its core, the relevance of systems theory to model, understand, and control synchronization is obvious and was recognized early, \eg{}~\cite{Nijmeijer:1997kr}. 

Synchronization is a pervasive concept in science and engineering. Currently, it is perhaps the most widely studied dynamical concept across systems biology~\cite{Winfree:1980ue,Glass:1988ub,Goldbeter:1996uo}, neuroscience~\cite{Hoppensteadt:1997tp,Izhikevich:2007vr}, chemistry~\cite{Kuramoto:1984wo}, physics~\cite{Huygens:1673vi,Pikovsky:2001et},  astronomy~\cite{Blekhman:1988wn},  and engineering~\cite{Strogatz:2003tm,Nijmeijer:2003uy}. Because synchronization involves interconnection at its core, the relevance of systems theory to model, understand, and control synchronization is obvious and was recognized early, \eg{}~\cite{Nijmeijer:1997kr}.

Two fundamental mathematical models of synchronization have emerged across the literature: the diffusive model and the kick model (a nickname throughout the paper for pulse-coupled synchronization model%
%\footnote{\textcolor{red}{Throughout the present paper, we adopt the names `kick model' (instead of `pulse-coupled model') and `kick synchronization', by opposition to `diffusive model' and `diffusive synchronization'.}}%
). The diffusive model analyzes synchronization as the result of diffusive coupling: the interconnection has the input--output interpretation of a static diffusive passive map. Owing to the fundamental homogenization nature of diffusion, diffusive interconnections tend to reduce differences between the time-course of interconnected variables, thereby favoring synchronized behavior if they are strong enough. In contrast, the kick model analyzes synchronization as the result of mutual rhythmic locking by short and weak pulses, akin to the physical phenomenon of resonance. 
The impulsive nature of the coupling combined with the continuous-time flow of the model between the pulses results in a hybrid model, see~\cite{Nunez:2012um} for a rigorous description of the kick model as a hybrid model.
%\textcolor{red}{It is indeed a hybrid model, that is, the composition of a continuous flow with successive discrete jumps induced by the pulses (see~\cite{Nunez:2012um} for a rigorous description of the kick model as a hybrid model).}

The diffusive model is largely dominant in the mathematical literature of synchronization. Synchronization between trajectories of state-space models is analyzed as an incremental stability property~\cite{Lohmiller:1998to} (\ie{}~the trajectories converge to one another rather than being attracted toward some equilibrium position). The leading concepts of Lyapunov analysis~\cite{Angeli:2002eb}, dissipativity analysis~\cite{Stan:2007jy,Arcak:2007dq,Sontag:2008tr,Scardovi:2010gx,Hamadeh:2011eu}, and---to a growing extent---contraction analysis~\cite{Pavlov:2004tb,Wang:2004jm,Slotine:2005vv,Pavlov:2005tu,Sontag:2010js,Russo:2010kx}, provide natural system theoretic tools to study synchronization. The literature of synchronization (closely related to consensus theory and coordination theory) is growing and the topic has attracted many systems and control researchers in the recent years. 

The kick model is largely dominant in natural manifestations of synchronization. Popular examples include synchronization of metronomes~\cite{YouTube:2011uh}, clocks~\cite{Huygens:1673tm,Bennett:2002uo}, heart beats~\cite{Peskin:1975wc}, flashing fireflies~\cite{Buck:1988ui}, neurons~\cite{Gerstner:2002ti}, earthquakes~\cite{Olami:1992tv}, and in fact most if not all spiking oscillators. In addition, kick synchronization is a source of inspiration for engineering applications (\eg{}~synchronization in wireless sensor networks~\cite{Hong:2005jw}, unsupervised classification problems~\cite{Rhouma:2001kl}). Despite the widespread occurrence of the phenomenon, %
% \textcolor{red}{and probably owing to the difficulties which characterize the theoretical analysis,}% 
the mathematical literature on kick synchronization is rather sparse compared to the literature on diffusive synchronization.

Primarily motivated by the recent thesis~\cite{Mauroy:2011th}, the present paper aims at comparing and contrasting the diffusive model and the kick model for the synchronization of periodic oscillators. We stress both the differences and the analogies between the two models, with a particular emphasis on their global stability properties. The discussion is tutorial in nature and focuses on simple examples, such as the coupling of van der Pol oscillators, which provides an insightful illustration of diffusive synchronization in the weakly nonlinear oscillation regime and of kick synchronization in the relaxation oscillation regime. The diffusive model is studied in continuous-time models while the kick model, hybrid in nature, is typically studied in discrete time. Ultimately, synchronization is always proven by showing that a certain distance between trajectories contracts over time, but the contraction measure is distinctively different in diffusive and kick models. 

While the diffusive and kick models of oscillator synchronization are fundamentally different, they also exhibit a remarkable analogy in the limit of weak coupling. This is because arbitrary oscillator models all reduce to one-dimensional phase models when the interconnection is sufficiently weak to maintain system trajectories in the neighborhood of the limit cycle oscillations of the uncoupled oscillators. Rooted in the seminal contributions of Winfree~\cite{Winfree:1967vf} and Kuramoto~\cite{Kuramoto:1975ki}, phase models of interconnected oscillators have a universal structure entirely characterized by their coupling function, which is strongly related to the phase response curve of the oscillators (\ie{}~a function which corresponds to the phase sensitivity of the uncoupled oscillators to an external perturbation). As a result, the fundamental difference between diffusive synchronization and kick synchronization is entirely coded in the shape of the coupling function, a phase map defined on the nonlinear unit circle. It is typically harmonic in the weak coupling limit of diffusive synchronization and typically monotone (and hence discontinuous) in the weak coupling limit of kick synchronization. Again, the synchronization mechanisms and the contraction measure are distinctively different even in the weak coupling limit, despite the shared model structure. 

\begin{figure}
	\centering
	\includegraphics{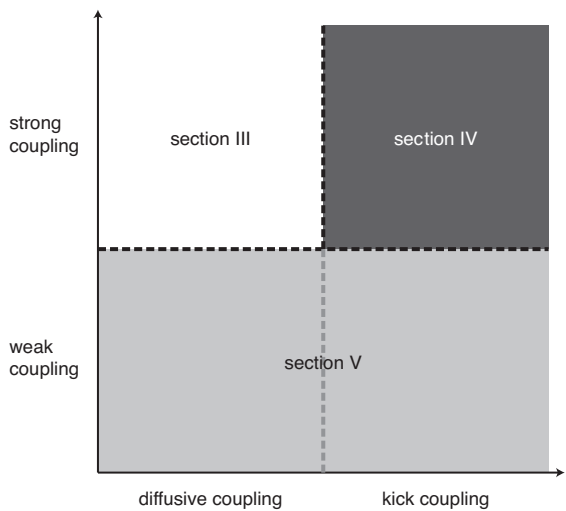}
	\caption{The paper is organized according to the coupling models. Sections~\ref{sec:diffusive} and~\ref{sec:kick} focus on (possibly strong) diffusive and kick synchronization, respectively. Section~\ref{sec:weak} deals with the phase models encountered in the limit of weak coupling.}
	\label{fig:content}
\end{figure}

The paper structure is illustrated in \myfigurename~\ref{fig:content}. Section~\ref{sec:models} reviews state-space models of oscillators and their phase reduction. The next sections present global stability results for the different synchronization models. Section~\ref{sec:diffusive} focuses on (possibly strong) diffusive synchronization while Section~\ref{sec:kick} focuses (possibly strong) kick synchronization. Section~\ref{sec:weak} deals with the phase models encountered in the limit of weak coupling. Section~\ref{sec:conclusion} provides concluding remarks.

%%%%%%% SECT.1 %%%%%%%
\section{Open oscillator models} \label{sec:models}

% Short introduction on state-space models vs phase models. Phase response curve. 
% Integrate-and-fire models. Infinite dimensional models. Kuramoto. 

% This section provides a short introduction to oscillators described as open $n$-dimensional state-space models and to their reduction to the one-dimensional (phase) center manifold of a hyperbolic periodic orbit. We first recall basic definitions about stable periodic orbits in dynamical systems (see~\cite{Guckenheimer:1983up,Farkas:1994uq} for details). We then introduce (finite and infinitesimal) phase response curves as fundamental mathematical information required for the reduction. We finally show how to reduce $n$-dimensional state-space models into one-dimensional phase models depending on the nature of the input.

This section provides a short introduction to oscillators viewed as open dynamical systems, that is, as dynamical systems that interact with their environment~\cite{Sepulchre:2006vk}.
We first recall basic definitions about stable periodic orbits in $n$-dimensional state-space models (see~\cite{Guckenheimer:1983up,Farkas:1994uq} for details).
We then introduce (finite and infinitesimal) phase response curves as fundamental mathematical information required for the reduction. 
We finally show how to reduce $n$-dimensional state-space models into one-dimensional phase models depending on the nature of the input.

\subsection{State-space models}

We consider open dynamical systems described by nonlinear time-invariant state-space models
\begin{subequations} \label{eq:nlsys}
	\begin{align} 
		\dot{\vecx} & = \vecF(\vecx) + \vecG(\vecx) u ,  	& \vecx & \in \mathbb{R}^n, u \in \mathbb{R}, \label{eq:nlsys_a} \\
		y & = \vecH(\vecx) , 									& y		  & \in \mathbb{R}, \label{eq:nlsys_b}
	\end{align}	
\end{subequations}
where the vector fields~$\vecF$ and~$\vecG$, and the measurement map~$\vecH$ support all usual smoothness conditions that are necessary for existence and uniqueness of solutions. We write $\vecPhi(t,\vecx_0,u)$ for the solution of the initial value problem~\eqref{eq:nlsys_a} with $\vecx(0)=\vecx_0$.

An oscillator is an open dynamical system whose zero-input steady-state behavior is periodic rather than constant. Formally, we assume that the zero-input system $\dot{\vecx} = \vecF(\vecx)$ admits a (locally hyperbolic) stable periodic orbit $\gamma$ with period~$T$ (and angular frequency $\omega=2\pi/T$). Picking an initial condition~$\vecx_0^\gamma$ on the periodic orbit~$\gamma$, this latter is described by the (nonconstant) periodic trajectory $\vecPhi(t,\vecx_0^\gamma,0)=\vecx^\gamma(t)$, such that $\vecx^\gamma(t) = \vecx^\gamma(t+T)$. The basin of attraction of $\gamma$ is the maximal open set $\mathcal{B}(\gamma)$ from which the periodic orbit attracts.

Since the periodic orbit $\gamma$ is homeomorphic to the unit circle~$\mathbb{S}^1$, it is naturally parametrized by a single scalar phase. Any point $\vecp\in\gamma$ is associated with a phase $\theta\in\mathbb{S}^1$, such that
\begin{equation*}
	\vecp = \vecx^\gamma(\theta/\omega)
\end{equation*}
(where $\vecx^\gamma_0$ is by convention associated with the phase $\theta = 0$).

For hyperbolic periodic orbits, the notion of phase is extended to any point $\vecq$ in the basin of attraction $\mathcal{B}(\gamma)$ through the concept of asymptotic phase. 
The asymptotic phase map $\Theta : \mathcal{B}(\gamma) \rightarrow \mathbb{S}^1$ assigns to each point~$\vecq$ in the basin $\mathcal{B}(\gamma)$ its asymptotic phase~$\theta\in\mathbb{S}^1$, such that 
\begin{equation} \label{eq:asymp-phase-map}
	\lim_{t\rightarrow+\infty} \left\| \vecPhi(t,\vecq,0) - \vecPhi(t,\vecx^\gamma(\theta/\omega),0) \right\|_2 = 0.
\end{equation}
This mapping is constructed such that the image of~$\vecx^\gamma_0$ is equal to $0$ and such that the progression along any orbit in $\mathcal{B}(\gamma)$ (in absence of perturbation) produces a constant increase in~$\theta$, that is, $\frac{d}{dt}\Theta(\vecPhi(t,\vecx_0,0)) = \omega$.

An isochron is a level set of the asymptotic phase map $\Theta$, that is, the set of all points in the basin of attraction of $\gamma$ characterized by a same asymptotic phase.

\begin{figure}
	\centering
	\includegraphics{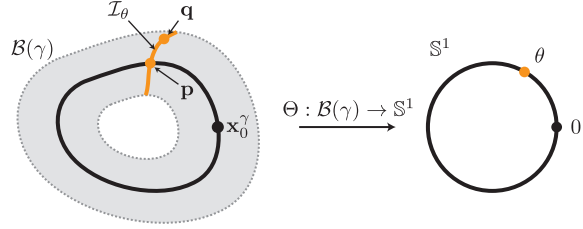}
	\caption{The asymptotic phase map $\Theta:\mathcal{B}(\gamma)\rightarrow\mathbb{S}^1$ assigns to each point~$\vecq$ in the basin $\mathcal{B}(\gamma)$ a single scalar phase $\theta$ on the unit circle $\mathbb{S}^1$, such that $\lim_{t\rightarrow+\infty}\left\| \vecPhi(t,\vecq,0) - \vecPhi(t,\vecp,0) \right\|_2=0$ where $\vecp=\vecx^\gamma(\theta/\omega)$. The set of all points $\vecq$ characterized by the same phase $\theta$ is the isochron~$\mathcal{I}_{\theta}$.}
	\label{fig:notations}
\end{figure}

\subsection*{The van der Pol oscillator: an illustrative model}

\begin{figure}
	\centering
	\includegraphics{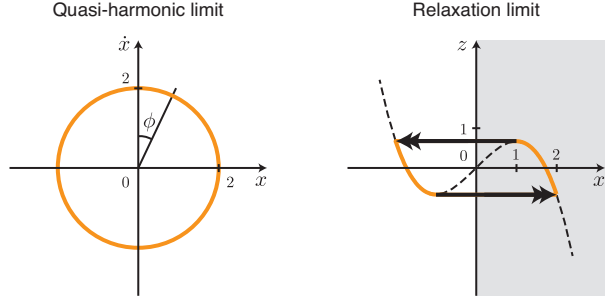}
	\caption{The van der Pol oscillator exhibits two different oscillation regimes: the quasi-harmonic ($\mu \ll 1$) and the relaxation ($\mu \gg 1$)  oscillation regimes. Quasi-harmonic and relaxation regimes are displayed in $(x,\dot{x})$ and $(x,z)$  state-spaces, respectively (with the transformation $z = x - x^3/3 - \dot{x}/\mu$).}
	\label{fig:vdp-regimes}
\end{figure}

In this paper, we illustrate most concepts on the van der Pol oscillator
\begin{equation} \label{eq:vanderpol}
	\ddot{x} - \mu (1-x^2) \dot{x} + x = \epsilon \bar{u}, \quad x\in\mathbb{R},
\end{equation}
where the parameter $\mu > 0$ measures the nonlinearity of the oscillator and the constant $\epsilon \geq 0$ measures the input strength (with $|\bar{u}|\leq 1$ for all times). Historically, this equation modeled a simple electrical circuit with nonlinear resistance and was used by van der Pol to study oscillations in vacuum tube circuits~\cite{VanderPol:1920um}. It played a seminal role in the development of nonlinear oscillation theory. One reason of this success is its ability (with only one parameter) to exhibit two very different regimes of oscillations (\myfigurename~\ref{fig:vdp-regimes}). For weak nonlinearities ($\mu \ll 1$), the oscillator displays quasi-harmonic oscillations. For strong nonlinearities ($\mu \gg 1$), it displays relaxation oscillations.

% This oscillator is able to generate both regimes: the quasi-harmonic limit for weak nonlinearity ($\mu \ll 1$) and the relaxation limit for strong nonlinearity ($\mu \gg 1$).

\subsubsection*{Quasi-harmonic limit}
To study the van der Pol oscillator in the quasi-harmonic limit, it is convenient to rewrite~\eqref{eq:vanderpol} in polar coordinates $(x,\dot{x}) = (r\sin(\phi),r\cos(\phi))$ as
\begin{subequations} \label{eq:vdp-polar}
	\begin{align}
		 \dot{r}    & = \mu g(r\sin(\phi),r\cos(\phi)) \cos(\phi) + \epsilon \cos(\phi) \bar{u} \label{eq:vdp-polar-r} \\
		 \dot{\phi} & = 1 - \frac{\mu}{r}  g(r\sin(\phi),r\cos(\phi)) \sin(\phi) - \frac{\epsilon}{r}\sin(\phi) \bar{u}  \label{eq:vdp-polar-phi}
	\end{align}
\end{subequations}
where we denote $g(x,\dot{x}) = (1-x^2)\dot{x}$ to simplify notations.

For small values of $\mu$ and $\epsilon$ ($\mu,\epsilon \ll 1$), standard averaging theory guarantees that $r$ stays in a $\mathcal{O}(\mu,\epsilon)$-neighborhood of $r^* = 2$ (see~\cite{Khalil:2002wj} for details). Substituting $r$ by $\tilde{r} = 2 + \mathcal{O}(\mu,\epsilon)$ into~\eqref{eq:vdp-polar-phi} and keeping first-order terms in the equation yield
\begin{equation} \label{eq:vdp-phi}
		\dot{\phi} \approx 1 - \frac{\mu}{2}  g(\tilde{r}\sin(\phi),\tilde{r}\cos(\phi)) \sin(\phi) - \frac{\epsilon}{2}\sin(\phi) \bar{u}.
\end{equation}
The approximately equal sign ($\approx$) means that~\eqref{eq:vdp-phi} neglects higher order terms in~$(\mu,\epsilon)$.
This one-dimensional equation describes the dynamics of the angular coordinate $\phi\in\mathbb{S}^1$. Note that this (geometrical) angular coordinate is different from the (temporal) asymptotic phase defined in~\eqref{eq:asymp-phase-map}. % \textcolor{red}{This difference comes from the nonconstant increase in $\phi$ along any orbit in absence of perturbation.}

\begin{figure}
	\centering
	\includegraphics{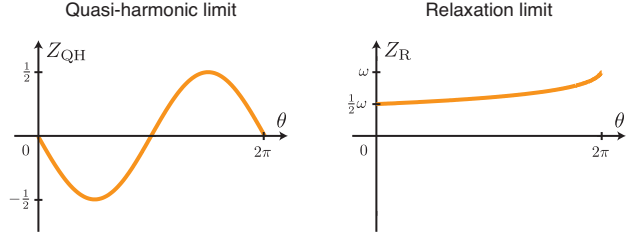}
	\caption{The shape of the infinitesimal phase response curve (for the van der Pol oscillator) is very different in both regimes. Typically, it is harmonic in the weakly nonlinear oscillation regime and monotone (and hence discontinuous) in the relaxation regime.}
	\label{fig:vdp-prc}
\end{figure}

Since the angular coordinate dynamics~\eqref{eq:vdp-phi} are one-dimensional, the asymptotic phase map appears as a bijective change of variable $\theta = \Theta(\phi)$ given by
\begin{equation*}
	\Theta(\phi) : \phi \mapsto \omega \int_0^\phi \frac{1}{1-\frac{\mu}{2}g(\tilde{r}\sin(\xi),\tilde{r}\cos(\xi))\sin(\xi)} d\xi.
\end{equation*}
This change of variable rescales the state-space and the (temporal) phase dynamics are given by
\begin{equation} \label{PRC_1}
	\dot{\theta} \approx \omega + \epsilon \underbrace{\frac{-\omega\sin(\phi)}{\tilde{r} - \mu  g(\tilde{r}\sin(\phi),\tilde{r}\cos(\phi)) \sin(\phi)}}_{\reveqdef Z_{\rm QH}(\theta)} \bar{u}
\end{equation}
where $\phi=\Theta^{-1}(\theta)$. The phase dynamics~\eqref{PRC_1} are the addition of two terms: the first term represents the autonomous angular frequency and the second term represents the influence of the input on the dynamics. The function~$Z_{\rm QH}(\cdot)$
captures the sensitivity of the oscillator phase dynamics to the input. It is known as the (input) infinitesimal phase response curve. (This notion will be defined properly in the next section.)
For values of $\mu$ tending to $0$, the asymptotic phase map~$\Theta$ tends to the identity, the angular frequency~$\omega$ tends to $1$, and the (input) infinitesimal phase response curve~$Z_{\rm QH}(\theta)$ tends to $-\frac{1}{2}\sin(\theta)$ (\myfigurename~\ref{fig:vdp-prc}).

\subsubsection*{Relaxation limit (and integrate-and-fire oscillators)}
To study the van der Pol oscillator in the relaxation limit, it is convenient to rewrite~\eqref{eq:vanderpol} in Li\'{e}nard's coordinates $(x,z)$ as
% \begin{subequations}
% 	\begin{align}
% 		\dot{x}     & = \mu \left( x - x^3/3 - z \right)  \\
% 		\mu \dot{z} & = x - \epsilon u
% 	\end{align}
% \end{subequations}
\begin{subequations} \label{eq:vdp-lienard}
	\begin{align}
		\frac{1}{\mu^2}x'     & = x - x^3/3 - z  \\
		z' & = x - \epsilon \bar{u}
	\end{align}
\end{subequations}
where we use the transformation $z = x - x^3/3 - \dot{x}/\mu$ and where $(\cdot)'$ denotes the derivative with respect to $s=t/\mu$.

For large values of $\mu$ ($1/\mu^2\ll1$), standard singular perturbation theory reduces the dynamics~\eqref{eq:vdp-lienard} to (see~\cite{Khalil:2002wj} for details)
\begin{align} \label{eq:vdp_IF}
	x' & = \frac{x}{1-x^2} -\epsilon\frac{1}{1-x^2} \bar{u}
\end{align}
on the critical manifold defined by $z=x-x^3/3$ and to instantaneous `jumps' at the folds in the critical manifold. Exploiting the central symmetry of the drift vector field (invariance under point reflection through the origin), we reduce the dynamics to the one-dimensional dynamics on the left branch of the critical manifold: the state~$x$ monotonically increases on $[-2,-1]$ according to~\eqref{eq:vdp_IF} and is reset to the lower threshold $\underline{x}=-2$ when reaching the upper threshold $\overline{x}=-1$.

Again, since the state dynamics are one-dimensional, there is a bijective change of variable $\theta = \Theta(x)$ given by
\begin{equation} \label{state_phase2}
	\Theta(x):x\mapsto \omega \int_{\underline{x}}^x \frac{1-\xi^2}{\xi}\, d\xi\,.
\end{equation}
This change of variable rescales in such a way that the lower threshold $\underline{x}=-2$ is mapped to $\theta=0$ and the upper threshold $\overline{x}=-1$ to $\theta=2\pi$.
The (temporal) phase dynamics are then given by 
\begin{equation}
\label{PRC_2}
	\theta' = \omega + \epsilon \underbrace{\left(-\frac{\omega}{x}\right)}_{\reveqdef Z_{\rm R}(\theta)}\bar{u}
\end{equation}
where $x=\Theta^{-1}(\theta)$. Here again, the phase dynamics~\eqref{PRC_2} are given by the addition of two terms: the autonomous angular frequency and the coupling term. In this case, the phase sensitivity function (or infinitesimal phase response curve)~$Z_{\rm R}(\cdot)$ is monotone on $[0,2\pi)$ (\myfigurename~\ref{fig:vdp-prc}).

In the relaxation limit, the van der Pol oscillator is equivalent to an integrate-and-fire model. More generally, the integrate-and-fire dynamics are  expressed as one-dimensional state dynamics between two threshold values~(see~\cite{Knight:1972hu,Abbott:1999ui}): a~scalar state variable $x$ monotonically increases between two thresholds $\underline{x}$ and $\overline{x}$, according to the dynamics
\begin{equation*}
	\dot{x} = \F(x), \quad \text{with $\F(x)>0$},
\end{equation*}
for all $x\in[\underline{x},\overline{x}]$. Upon reaching the upper threshold $\overline{x}$, the state is instantaneously reset to the lower threshold $\underline{x}$. Roughly speaking, the oscillator integrates between the two thresholds and fires when reaching the upper threshold.

The most popular integrate-and-fire oscillator is the leaky integrate-and-fire (LIF) oscillator, characterized by the monotone vector field $\F(x)=S+R x>0$, \mbox{$\forall x\in[\underline{x},\overline{x}]=[0,1]$}. An important generalization of the LIF oscillator---in the sense that the dynamics are not monotone anymore---is the quadratic integrate-and-fire (QIF) oscillator, defined by the vector field \mbox{$\F(x)=S+x^2$}, with $S>0$~\cite{Ermentrout:1986ve}.

Similarly to~\eqref{state_phase2}, the asymptotic phase map $\Theta$ that corresponds to the integrate-and-fire dynamics is the bijective change of variable given by
\begin{equation*} %\label{state_phase}
\Theta(x):x\mapsto \omega \int_{\underline{x}}^x \frac{1}{\F(\xi)}\, d\xi\,,
\end{equation*}
with the lower threshold $\underline{x}$ (resp. the upper threshold $\overline{x}$) being mapped to $\theta=0$ (resp. $\theta=2\pi$).

It is worth mentioning that, in the relaxation limit, the van der Pol oscillator model closely resemble the popular model of FitzHugh-Nagumo~\cite{Fitzhugh:1961il,Nagumo:1962iz}, a two-dimensional qualitative reduction of Hodgkin-Huxley model of neuronal action potentials~\cite{Hodgkin:1952td}. Integrate-and-fire models are broadly used in neurodynamics~\cite{Knight:1972hu,Abbott:1999ui}.

\subsection{Phase response curves} \label{subsec_PRC}

For many oscillators, the structure of the asymptotic phase map and therefore the topology of isochrons are very complex. This often makes their analytical computation impossible and even their numerical computation intractable (at least very expensive for high dimensional oscillator models), an issue that prevents from building an exact one-dimensional phase model valid in the whole basin of attraction. However, in many situations, a complete knowledge of the isochrons is not required to study the oscillator dynamics. Instead, it is sufficient to consider the phase response curve, as it has naturally appeared in~\eqref{PRC_1} and~\eqref{PRC_2} through the reduction of the van der Pol dynamics.

Starting with the pioneering work of Winfree~\cite{Winfree:1967vf,Winfree:1980ue}, the phase response curve of an oscillator has proven a useful input--output tool to study oscillator dynamics. It indicates how the timing of inputs affects the timing (steady-state phase shift) of oscillators. Phase response curves are directly related to isochrons but capture only partial information about them.

\begin{definition}
	The finite \emph{Phase Response Curve (PRC)} corresponding to a Dirac delta input $u(\cdot) = \epsilon \delta(\cdot)$ is the map $Z_{\epsilon}:\mathbb{S}^1\rightarrow(-\pi,\pi]$ defined as
	\begin{equation*}
		Z_\epsilon(\theta) = \lim_{t\rightarrow 0^+}\underbrace{\Theta( \vecPhi(t,\vecx^\gamma(\theta/\omega),\epsilon \delta(\cdot)) )}_{\text{post-stimulus phase}}  -  \underbrace{\Theta( \vecPhi(t,\vecx^\gamma(\theta/\omega),0) )}_{\text{pre-stimulus phase}}.
	\end{equation*}
	It associates with each point on periodic orbit (parametrized by its phase $\theta$) the phase shift induced by the input.
\end{definition}

In many situations, the PRC can be determined experimentally. Moreover, it can be obtained numerically by computing the perturbed and unperturbed trajectories of the nonlinear state-space model and by comparing the asymptotic phase difference between each pair of trajectories.

A mathematically more abstract---yet very useful---tool is the infinitesimal phase response curve, which appears in~\eqref{PRC_1} and~\eqref{PRC_2}. It records essentially the same information as the finite phase response curve but for infinitesimally small Dirac delta input ($\epsilon \ll 1$). 
\begin{definition}
	The \emph{(input) infinitesimal Phase Response Curve (iPRC)} is the map $Z:\mathbb{S}^1\rightarrow\mathbb{R}$ defined as the directional derivative 
	\begin{equation*}
		Z(\theta) = D\Theta(\vecx^\gamma(\theta/\omega))[\vecG(\vecx^\gamma(\theta/\omega))]
	\end{equation*}
	where
	\begin{equation*}
		D\Theta(\vecx)[\vec{\eta}] = \lim_{\epsilon\rightarrow0}\frac{\Theta(\vecx+\epsilon\vec{\eta}) - \Theta(\vecx)}{\epsilon}.
	\end{equation*}
	The directional derivative can be computed as the inner product
	\begin{equation*}
		D\Theta(\vecx)[\vecG(\vecx)] = \langle \nabla_{\vecx}\Theta(\vecx) , \vecG(\vecx) \rangle
	\end{equation*}
	where $\nabla_{\vecx}\Theta(\vecx)$ is the gradient of $\Theta$ at $\vecx$ and is known as the state infinitesimal phase response curve.
\end{definition}

\begin{remark} \label{PRC_weak_coupling}
	For small values of $\epsilon$ ($\epsilon \ll 1$), the finite phase response curve is well approximated by the infinitesimal phase response curve, that is $Z_\epsilon(\cdot) \approx \epsilon Z(\cdot)$. \hfill $\lrcorner$
\end{remark}

\begin{figure}
	\centering
	\includegraphics{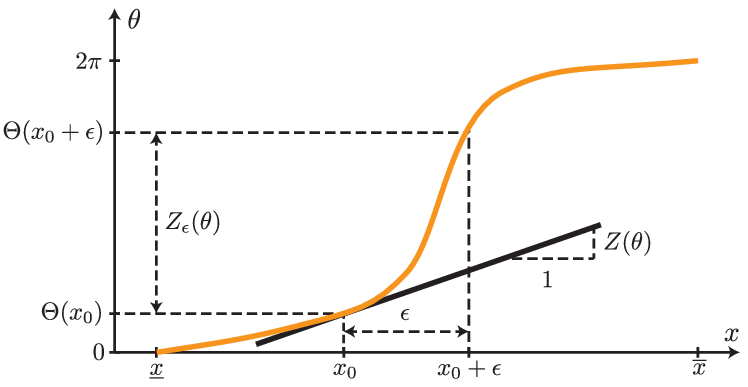}
	\caption{For integrate-and-fire oscillators, the finite PRC $Z_\epsilon$ is directly derived from the iPRC~$Z$.}
	\label{PRC_impulsive}
\end{figure}

\begin{remark}[Integrate-and-fire oscillators]
	For integrate-and-fire oscillators, the iPRC has the exact analytical expression (see~\cite{Brown:2004iy,Izhikevich:2007vr})
	\begin{equation}
		Z(\theta)=\frac{\omega}{\F(x^\gamma(\theta/\omega))}\,.
	\end{equation}
	Moreover, the finite PRC is directly obtained from the iPRC due to the unidimensional nature of the state-space. Namely, it follows from \myfigurename~\ref{PRC_impulsive} that
	\begin{equation*}
		Z_\epsilon(\theta)=\int_{x^\gamma(\theta/\omega)}^{x^\gamma(\theta/\omega)+\epsilon} Z(\Theta(\xi))\,d\xi \,.
	\end{equation*}
	For LIF oscillators, the iPRC and the finite PRC are monotone, as for the relaxation van der Pol model (\myfigurename~\ref{fig:vdp-prc}). \hfill $\lrcorner$
\end{remark}

\subsection{Reduced phase models} \label{subsec_couplmodel}

We review two popular phase models, which are obtained through phase reduction methods in the case of weak input and impulsive input, respectively~\cite{Kuramoto:1984wo,Kuramoto:1997kd,Hoppensteadt:1997tp,Brown:2004iy,Izhikevich:2007vr}.

\subsubsection{Weak input}
In the weak perturbation limit, that is, for small input
\begin{equation*}
	u(t) = \epsilon \bar{u}(t), \quad \epsilon \ll 1, \quad |\bar{u}(t)|\leq1 \text{ for all $t$}, 
\end{equation*}
any solution $\vecPhi(t,\vecx_0,u)$ of the oscillator model which starts in the neighborhood of the hyperbolic stable periodic orbit~$\gamma$ stays in its neighborhood. The $n$-dimensional state-space model can then be approximated by a one-dimensional continuous-time phase model
\begin{equation} \label{eq:weak_coupling}
	\dot{\theta} = \omega + \epsilon Z(\theta) \bar{u}(t)
\end{equation}
where the phase variable $\theta$ evolves on the unit circle $\mathbb{S}^1$. The phase model is fully characterized by the angular frequency $\omega>0$ and by the iPRC $Z:\mathbb{S}^1 \rightarrow \mathbb{R}$.

\subsubsection{Impulsive input (kick)} 
In the impulsive perturbation limit, the input corresponds to delta-like kicks of amplitude~$\epsilon$ (not necessarily small), that is,
\begin{equation*}
	u(t) = \epsilon \sum_{k=0}^{\infty} \delta(t-t_{k})\,.
\end{equation*}
Any solution $\vecPhi(t,\vecx_0,u)$ of the oscillator model which starts from the periodic orbit $\gamma$ leaves the periodic orbit under the effect of a kick and converges back to the periodic orbit. If the periodic orbit is sufficiently strongly attractive, the trajectory will be back in the neighborhood of the periodic orbit before the next kick takes place. The $n$-dimensional state-space model can then be approximated by a one-dimensional hybrid phase model, with
\begin{subequations} \label{equa_hybrid}
	\begin{enumerate}
		\item the (constant-time) flow rule 
		\begin{align}
			\dot{\theta} & = \omega, & \text{for all $t \neq t_k$}, \label{equa_hybrid1} \\
		\intertext{\item and the (discrete-time) jump rule (\ie{}~the kick)}
			\theta^+  & = \theta + Z_{\epsilon}(\theta), & \text{for all $t = t_k$}, \label{equa_hybrid2}
		\end{align}
	\end{enumerate}	
\end{subequations}
where the phase variable $\theta$ evolves on the unit circle $\mathbb{S}^1$. The phase model is fully characterized by the angular frequency $\omega>0$ and by the PRC $Z_\epsilon:\mathbb{S}^1\rightarrow(-\pi,\pi]$.

% \begin{equation}
% 	\operatorname{PTC}_{u}:\mathbb{S}^1 \rightarrow \mathbb{S}^1
% \end{equation}
% \begin{equation}
% 	\operatorname{PRC}_{u}:\mathbb{S}^1 \rightarrow (-\pi,\pi]
% \end{equation}

%%%%%%% SECT.2 %%%%%%%
\section{Diffusive synchronization} \label{sec:diffusive}

% Connecting two van der Pol oscillators with a wire. 
% Output feedback incremental passivity. Literature.
% Strong coupling.

\subsection{Connecting two van der Pol oscillators with a resistor}

\begin{figure}
	\centering
	\includegraphics{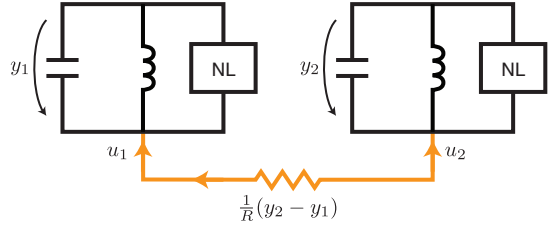}
	\caption{The interconnection of two van der Pol oscillators with a resistor causes a current flow proportional to the voltage difference $(y_2-y_1)$. This interconnection is known as diffusive coupling, a name which comes from the diffusive nature of the resistor.}
	\label{fig:coupled-vdp}
\end{figure}

Diffusive synchronization is a model of physical interconnection through a diffusive medium. As a simple illustration of diffusive synchronization, we consider two van der Pol oscillators interconnected with a resistor (\myfigurename~\ref{fig:coupled-vdp})
\begin{subequations} \label{eq:vdp-diff}
	\begin{align}
		\dot{x}_i & = - w_i + \mu(x_i - x_i^3/3)  + u_i \\
		\dot{w}_i & = x_i \\
		y_i &= x_i 
	\end{align}
\end{subequations}
where $x_i$ and $w_i$ denote the voltage across the capacitor and the current through the inductor, respectively. 

The interconnection with a resistor induces a current flow proportional to the voltage difference $(y_2 - y_1)$ and inversely proportional to the resistance $R$. The smaller the resistance~$R$, the higher the coupling strength~$K = 1/R$. Using the vector notation $\vec{u} = \tr{(u_1,u_2)}$ and $\vec{y} = \tr{(y_1,y_2)}$, the interconnection is expressed as
\begin{equation*}
	\vec{u} = - L \vec{y}
\end{equation*}
with the coupling matrix given by
\begin{equation} \label{eq:vdp-diff-matrix}
	L = K 
	\begin{bmatrix}
		\phantom{-}1 & -1 \\
		-1 & \phantom{-}1
	\end{bmatrix} .
\end{equation}
This type of coupling is known as diffusive coupling, owing to the nature of the resistor.

In this context, synchronization is a convergence property for the difference between the solutions of different systems. 	Suppose that we have a network of $N$ oscillators. The oscillators are said to output synchronize if
\begin{equation*}
	\lim_{t\rightarrow+\infty}\| y_i(t) - y_j(t) \| = 0, \quad \forall i,j = 1,\ldots,N
\end{equation*}
where $\|\cdot\|$ denotes the Euclidean norm of the enclosed signal.

In the following, we show under which conditions the synchronization is guaranteed using: incremental stability theory, incremental passivity theory, and contraction theory.

\subsubsection{Incremental stability theory}
Convergence properties for the difference between solutions of a closed system are characterized by notions of incremental stability~\cite{Angeli:2002eb}. 

Considering the error variable $\vece = \tr{(e_x,e_w)} = \vecx_1 - \vecx_2$, the error system is written as follows
\begin{equation*}
	\dot{\vece} = 
	\begin{bmatrix}
		-2K+\mu & -1 \\
		1 & \phantom{-}0 \\
	\end{bmatrix}
	\vece  -
	\begin{bmatrix}
		\Delta\phi(x) \\
		0 \\
	\end{bmatrix} 
\end{equation*}
where $\phi(\cdot)$ stands for the monotonic function $\phi(s) = \mu \, s^3/3$ and  $\Delta\phi(\cdot)$ is defined as $\Delta \phi(s) = \phi(s_1) - \phi(s_2)$.
The monotonicity property of the nonlinearity $\phi(\cdot)$ implies that 
\begin{equation*}
	[\phi(s_1) - \phi(s_2)] \, (s_1 - s_2) = \Delta\phi(s) \, \Delta s \geq 0
\end{equation*}
for all $\Delta s = s_1 -s_2$.
Considering the Lyapunov function 
\begin{equation*}
	V = \frac{1}{2}(e_x^2 + e_y^2) \geq 0,
\end{equation*}
we obtain
\begin{align*}
	\dot{V} & = (\mu -2K)\,e_x^2 - \underbrace{\Delta \phi(x) \, e_x}_{\geq 0} \leq (\mu -2K)\,e_x^2.
\end{align*}
Then by the Lyapunov stability theorem and the invariance principle~\cite[Theorem 4.1 and 4.4]{Khalil:2002wj}, we conclude to the synchronization of both oscillators for $2K > \mu$.

\subsubsection{Incremental passivity theory}
For open systems, the  notion corresponding to incremental stability is incremental dissipativity~\cite{Stan:2007jy}. 
Denoting the incremental variables by $\Delta\vecx = \tr{(\Delta x_x,\Delta x_w)} = \vecx_1 - \vecx_2$, $\Delta u = u_1 - u_2$, and $\Delta y = y_1 - y_2$, a system is incrementally passive if it satisfies a dissipative inequality
\begin{equation*}
	\dot{\Delta S} \leq w(\Delta u,\Delta y)
\end{equation*}
for an incremental scalar storage function $\Delta S(\Delta \vecx) \geq 0$ with a supply rate $w(\Delta u,\Delta y)$.
 
The incremental system of~\eqref{eq:vdp-diff} is written as follows
\begin{subequations}
	\begin{align*}
		\dot{\Delta \vecx} & = 
		\begin{bmatrix}
			\mu  & -1 \\
			1 & \phantom{-}0 \\
		\end{bmatrix}
		\Delta\vecx + 
		\begin{bmatrix}
			1 \\ 0
		\end{bmatrix}
		\Delta u - 
		\begin{bmatrix}
			\Delta\phi(x) \\ 0
		\end{bmatrix} \\
		\Delta y  & = 
		\begin{bmatrix}
			1 & 0
		\end{bmatrix}
		\Delta \vecx .
	\end{align*}
\end{subequations}
Considering the incremental storage 
\begin{equation*}
	\Delta S(\Delta \vecx) = \frac{1}{2}(\Delta x_x^2 + \Delta x_w^2) \geq 0,
\end{equation*}
we have
\begin{align*}
	\dot{\Delta S} & = \mu \,\Delta y^2 - \underbrace{\Delta y \, \Delta\phi(y)}_{\geq 0} + \Delta u \,\Delta y  \leq \mu \,\Delta y^2 + \Delta u\, \Delta y .
\end{align*}
Substituting $\Delta u = - 2K \Delta y$ in the previous equation yields
\begin{equation*}
	\dot{\Delta S} \leq ( \mu - 2K) \,\Delta y^2,
\end{equation*}
which implies asymptotic convergence of $\Delta y$ to zero (that is, output synchronization) when $2K>\mu$.

\subsubsection{Contraction theory} 

Nonlinear contraction theory gives a simple yet general method to study synchronization~\cite{Wang:2004jm}. If the dynamics equations verify
\begin{equation*} 
	\dot{\vecx}_1 - \vech(\vecx_1,t)  = \dot{\vecx}_2 - \vech(\vecx_2,t)
\end{equation*}
where the function $\vech$ is contracting, then $\vecx_1$ and $\vecx_2$ will converge to each other exponentially.

Considering the following vector field
\begin{equation*}
	\vech(\vecx,t) = 
	\begin{bmatrix}
		-w + \mu\,(x - x^3/3) - 2K\,x \\
		x
	\end{bmatrix} ,
\end{equation*}
the Jacobian matrix is given by
\begin{equation*}
	J = 
	\begin{bmatrix}
	  	(\mu - 2K) - \mu \, x^2  & -1 \\
		1 &0
	\end{bmatrix}
\end{equation*}
and is negative semidefinite for $2K>\mu$. This implies that $\vech$ is contracting and that both oscillators synchronize when $2K>\mu$.

\subsection{Large networks}

Most collective phenomena among oscillators in nature arise in large networks of oscillators. The notion of diffusive coupling and the tools described in the previous section can be extended to a network of $N$ oscillators.

Each oscillator dynamics is written, for $i=1,\ldots,N$, as
\begin{subequations} \label{eq:iooscill}
	\begin{align}
		\dot{\vecx}_i & = \vecF(\vecx_i) + \vecG(\vecx_i) u_i ,  	\\
		y_i & = \vecH(\vecx_i) . 										  
	\end{align}
\end{subequations}
The general diffusive interconnection is then given by
\begin{equation} \label{eq:diffusive_coupling}
	u_i = \sum_{j\in\mathcal{N}_i} K_{ji}(y_{j} - y_{i}), \quad i = 1,\ldots,N
\end{equation}
where $K_{ji}$ is a positive constant and $\mathcal{N}_i\subseteq\mathcal{N}$ is the subset of oscillators transmitting their outputs to the $i$th~oscillator. (The set $\mathcal{N} = \{1,\ldots,N\}$ denotes all oscillators in the network.)
Using the vector notations $\vec{u} = \tr{(u_1,\ldots,u_N)}$ and $\vec{y} = \tr{(y_1,\ldots,y_N)}$, the interconnection is expressed as $\vec{u} = - L \vec{y}$ with the coupling matrix $L$ defined as the Laplacian of the network graph
\begin{equation*} 
	L_{ij} =
	\begin{cases}
		\sum_{j\in\mathcal{N}_i\backslash\{i\}} K_{ji} & \text{if $i=j$}, \\
		- K_{ji} & \text{if $j\in\mathcal{N}_i\backslash\{i\}$}, \\
		0 & \text{otherwise} .
	\end{cases}
\end{equation*}

The diffusive coupling $\vec{u} = - L \vec{y}$ is a passive operator. Neglecting the symmetry neutral mode $L\mathbf{1} = 0$, the excess of passivity of the operator is given by the smallest nonzero eigenvalue of the symmetric part of $L$, which is the parameter $2K$ in~\eqref{eq:vdp-diff-matrix}.

Synchronization is guaranteed in a network of input--output oscillators~\eqref{eq:iooscill} if the excess of passivity in the coupling ($2K$ in the example~\eqref{eq:vdp-diff-matrix}) compensates for the shortage of incremental passivity of the model~\eqref{eq:iooscill} ($\mu$ in van der Pol example). A precise statement of this result is found in~\cite{Stan:2007jy} under the assumption of a balanced graph, \ie{}~when $L + \tr{L} \geq 0$. The balancing assumption has been elegantly removed in the recent paper~\cite{Chopra:2012wo}. 

A limitation of diffusive coupling model is that the incremental stability analysis often suggests the necessity of a strong enough coupling. This is in contrast to many synchronization problems in which coupling strengths are weak. 
% The study of collective phenomena arising from those weak interconnections is then questionable \textcolor{red}{in this context. Therefore, this issue could motivate future research to develop a general framework which is better suited to the study of diffusive synchronization ??mention differential Lyapunov framework (cf paper Fulvio-Rodolphe)??}

%%%%%%% SECT.3 %%%%%%%
\section{Kick synchronization} \label{sec:kick}

\subsection{Connecting two van der Pol oscillators with impulses}

\begin{figure}
	\centering
	\includegraphics[width=6cm]{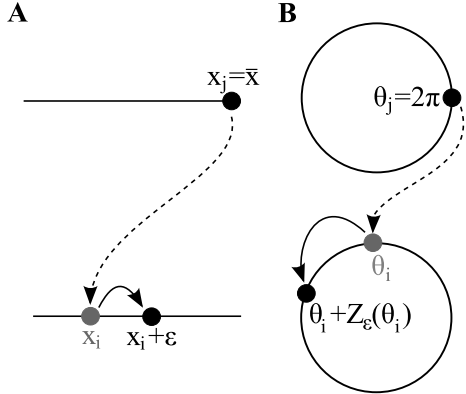}
	\caption{A. Impulsive coupling with integrate-and-fire oscillators: an oscillator reaching the upper threshold $x=\overline{x}$ triggers an instantaneous increment $\epsilon$ to the state of the other oscillators. B. Impulsive coupling with phase oscillators: an oscillators reaching $\theta=2\pi$ triggers an instantaneous increment $Z_\epsilon(\theta)$ to the phase of the other oscillators.}
	\label{impulsive_coupling}
\end{figure}

A mathematical model of kick synchronization of (identical) oscillators connected through impulsive coupling was first proposed by Peskin, in the particular case of integrate-and-fire oscillators~\cite{Peskin:1975wc}. When the oscillators fire, they send out a kick that causes an instantaneous increment~$\epsilon$ to the state of all other oscillators of the network (\myfigurename~\ref{impulsive_coupling}A).

\begin{figure}
	\centering
	\includegraphics[width=8.5cm]{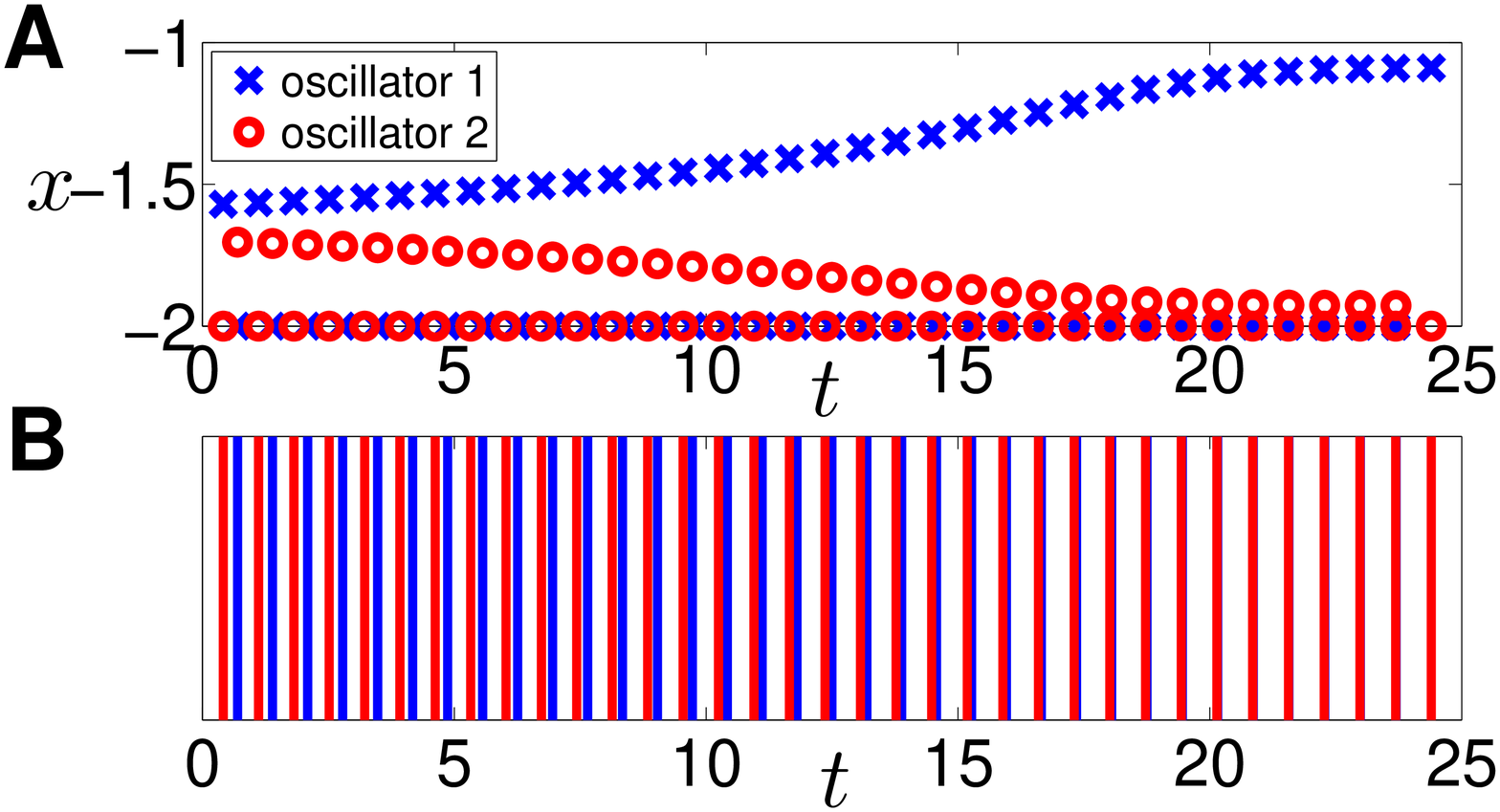}
	\caption{Synchronization of van der Pol oscillators with kick coupling ($\epsilon>0$). A. The instantaneous state values of an oscillator at the successive firings of the other oscillator approach either the lower threshold ($\underline{x}=-2$) or the upper threshold ($\overline{x}=-1$). (The blue and red symbols represent the instantaneous state values of the oscillators at the successive firing times.) B. After a short transient, the oscillators fire in unison. (The blue and red lines represent the firing times $t_k$ of oscillator 1 and 2, respectively.)}
	\label{impulsive_vdp_synchro}
\end{figure}

\begin{figure}
	\centering
	\includegraphics[width=8.5cm]{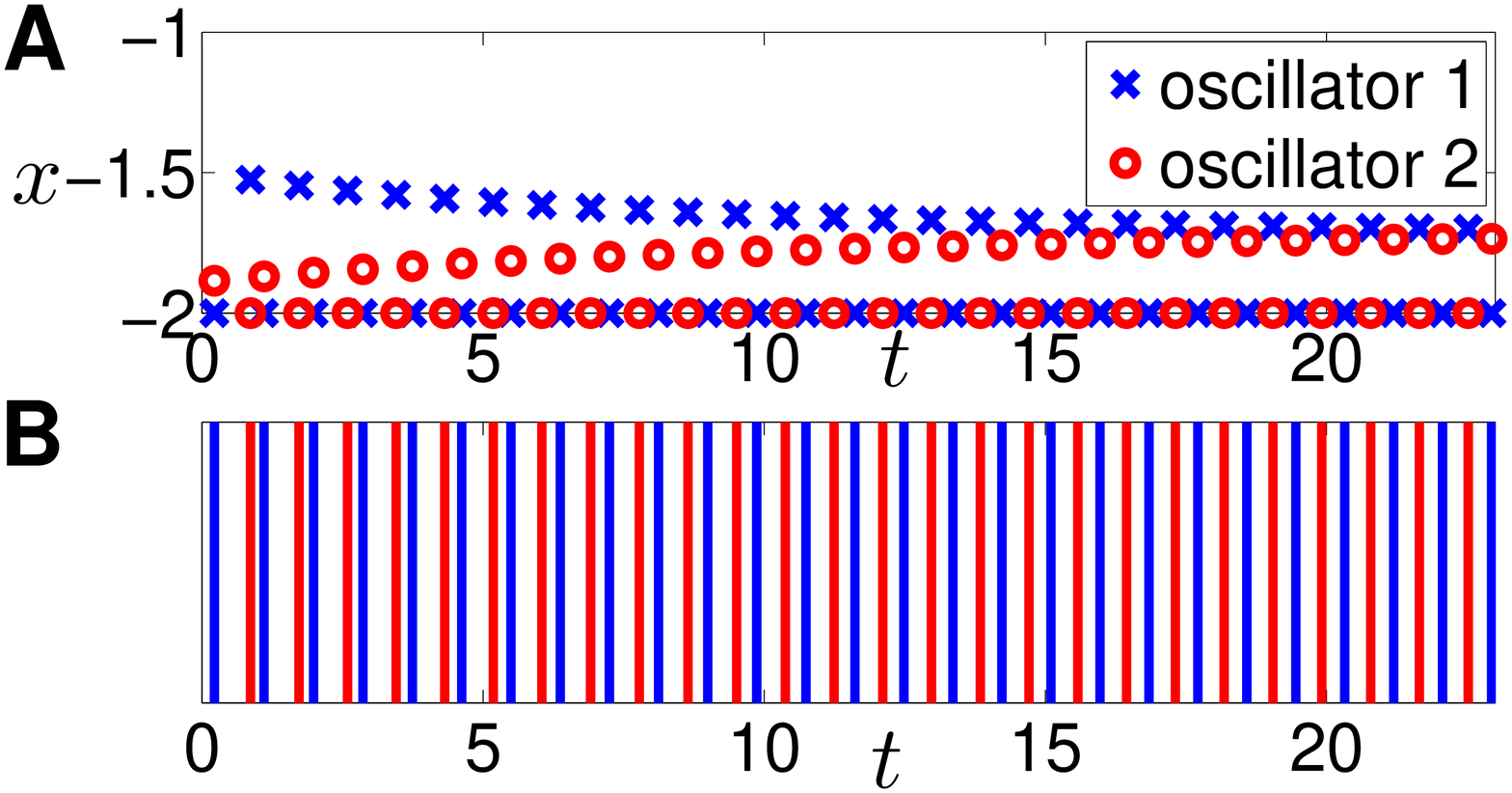}
	\caption{Phase-locking of van der Pol oscillators with kick coupling ($\epsilon<0$). A. The instantaneous state values of an oscillator at the successive firings of the other oscillator asymptotically converge to a constant value $x\approx -1.7$. (The blue and red symbols represent the instantaneous state values of the oscillators at the successive firing times.) B. The asymptotic firing pattern of the oscillators is periodic. (The blue and red lines represent the firing times $t_k$ of oscillator~1 and 2, respectively.)}
	\label{impulsive_vdp_phaselocked}
\end{figure}

As a first illustration of Peskin's impulsive coupling, consider two van der Pol oscillators in the relaxation limit \mbox{$\mu \gg 1$.} In good approximation, the oscillators are characterized by the integrate-and-fire dynamics~\eqref{eq:vdp_IF}, with \mbox{$x\in[-2,-1]$.} In addition, suppose that an oscillator which fires (\ie{}~which reaches $x=-1$) at time $t_k$ sends out a kick $u(t)=\epsilon \delta(t-t_k)$ which increases the state $x$ of the other oscillator by a value $\epsilon/(x^2-1)$, according to~\eqref{eq:vdp_IF}. Through a well-chosen change of variable, this increment can be made constant for any state value, so that the model is equivalent to Peskin model. Numerical simulations show that the two oscillators achieve synchronization: after a short transient period, they fire in unison (\myfigurename~\ref{impulsive_vdp_synchro}). When a negative increment $\epsilon<0$ is considered, the oscillators asymptotically converge to a phase-locked configuration: they fire at a constant rate and they are characterized by the same instantaneous state values (\ie{}~$x=-2$ for the firing oscillator and $x\approx -1.7$ for the other oscillator) at each firing time $t_k$ (\myfigurename~\ref{impulsive_vdp_phaselocked}).

In this paper, we use the more general (but equivalent) definition of Peskin's impulsive coupling in terms of phase, a definition which is motivated by the developments of Sections~\ref{subsec_PRC} and~\ref{subsec_couplmodel} (\myfigurename~\ref{impulsive_coupling}B). Similarly to~\eqref{equa_hybrid}, each phase oscillator obeys
% \begin{enumerate}
% 	\item the (constant) flow rule 
% 	\begin{equation}
% 		\label{flow_rule}\ie{}~
% 		\dot{\theta}_i=\omega\,;
% 	\end{equation}
% 	\item and the jump rule (\ie{}~the kick)
% 	\begin{equation}
% 		\label{jump_rule}
% 		\begin{array}{l}
% 			\textrm{if } \exists j:\theta_j=2\pi \\
% 			\textrm{  then  } \theta_i^+=\min\{\theta_i+Z_\epsilon(\theta_i),2\pi\} \quad \forall i\neq j\,.
% 		\end{array}
% 	\end{equation}
% \end{enumerate}
\begin{subequations} \label{rule}
	\begin{enumerate}
		\item the (constant-time) flow rule 
		\begin{align}
			\dot{\theta}_i & =\omega, & \text{if $\forall j\neq i : \theta_j \neq 2\pi$}, \label{flow_rule} \\
		\intertext{\item and the (discrete-time) jump rule (\ie{}~the kick)}
			\theta_i^+     & =\min\{\theta_i+Z_\epsilon(\theta_i),2\pi\}, & \text{if $\exists j\neq i :\theta_j = 2\pi$}, \label{jump_rule}
		\end{align}
	\end{enumerate}
\end{subequations}
(Note that the coupling is all-to-all, \ie{}~$\mathcal{N}_i=\mathcal{N}\setminus\{i\}$.) The threshold imposed in~\eqref{jump_rule} corresponds to the absorption phenomenon. If the kick is strong enough, an oscillator may aggregate with the oscillator that triggered the kick. The two oscillators have subsequently the same phase and create a cluster which behaves as a single oscillator. (We therefore make no distinction between a single oscillator and a cluster.)

\begin{remark}
	The jump rule~\eqref{jump_rule} corresponds to an excitatory coupling, that is $Z_\epsilon(\theta)>0$ $\forall \theta\in(0,2\pi)$. For the sake of simplicity, we adopt this assumption in the sequel. The extension of the results to non-excitatory couplings is straightforward. \hfill $\lrcorner$
\end{remark}

\subsection{Large networks and firing maps}

The mathematical analysis of kick synchronization differs from diffusive synchronization in that it can be achieved through the analysis of a discrete-time model. Since the network is uncoupled between two kicks, all the information is retained by considering the network state at the discrete kick times only. For instance, if the configuration of two coupled oscillators right after a kick is $(\theta,2\pi)$, then their configuration right after the next kick is given by $(2\pi,\firemap(\theta))$, with
\begin{equation*}
	\firemap(\theta)=2\pi-\theta+Z_\epsilon(2\pi-\theta)\,.
\end{equation*}
The discrete-time map $\theta^+=\firemap(\theta)$ expresses the phase differences between the two oscillators at the successive kick times. It was originally introduced in~\cite{Mirollo:1990ft} as the so-called firing map.

For the study of large networks, we first assume that the order of the oscillators is not modified under the effect of~\eqref{jump_rule}, that is we assume that $\theta_1+Z_\epsilon(\theta_1)<\theta_2+Z_\epsilon(\theta_2)$ if $\theta_1<\theta_2$. This assumption, which is always satisfied for one-dimensional oscillators such as integrate-and-fire oscillators, is summarized as follows:
\begin{assumption}[Order preserving assumption]
	\label{order_preserv}
	The finite phase response curve satisfies the condition $Z_\epsilon'(\theta)>-1$ $\forall \theta\in(0,2\pi)$. \hfill $\lrcorner$
\end{assumption}

Provided that Assumption~\ref{order_preserv} holds, the snapshot configurations of a network of $N$ oscillators are given by the successive iterations of a $(N-1)$-dimensional firing map, which appears as a straightforward generalization of the scalar firing map:
\begin{equation}
	\label{N_fir_map}
	\Firemap[(\theta_1,\dots,\theta_{N-1})] = 
	\left\{ 
	\begin{array}{l} 
		\firemap(\theta_{N-1}) \\
		\firemap(\theta_{N-1}-\theta_1) \\
		% \hspace{0.7cm} \vdots \\
		\phantom{\firemap(\theta_{N-1}} \vdots \\
		\firemap(\theta_{N-1}-\theta_{N-2})
	\end{array} 
	\right.\,.
\end{equation}
Note that the oscillators are not assigned constant indices but are labeled at each kick according to the phase ordering $0<\theta_1<\theta_2<\cdots<\theta_{N-1}<\theta_N=2\pi$.

An extensive study of kick synchronization in large networks is therefore restricted to the (global) stability analysis of the firing map~\eqref{N_fir_map}. The strongest stability result is obtained for oscillators characterized by a monotone PRC~$Z_\epsilon$~\cite{Mauroy:2008dp}.
\begin{theorem}
	\label{theo_fir_map}
	Consider a finite PRC that satisfies (i) Assumption~\ref{order_preserv} and (ii) either $Z''_\epsilon(\theta)>0$ $\forall \theta\in(0,2\pi)$ or $Z''_\epsilon(\theta)<0$ $\forall \theta\in(0,2\pi)$. Then, the $(N-1)$-dimensional firing map~\eqref{N_fir_map}, with $N>1$, has a contraction property with respect to the $1$-norm
	\begin{equation}
		\label{1_norm}
		\|(\theta_1,\cdots,\theta_{N-1})\|=|\theta_1|+\sum_{k=1}^{N-2}|\theta_k-\theta_{k+1}|+|\theta_{N-1}|\,.
	\end{equation}
	That is,
	\begin{itemize}
		\item the firing map is contracting with respect to~\eqref{1_norm} if $Z'_\epsilon(\theta)<0$ $\forall \theta\in(0,2\pi)$;
		\item the firing map is expanding with respect to~\eqref{1_norm} if $Z'_\epsilon(\theta)>0$ $\forall \theta\in(0,2\pi)$. \hfill $\lrcorner$
	\end{itemize}
\end{theorem}

A straightforward corollary of Theorem~\ref{theo_fir_map} is that kick synchronization has two important features: (i) \emph{isolated phase-locked configuration} and (ii) \emph{finite-time synchronization}.

\begin{figure}
	\centering
	\includegraphics[width=7cm]{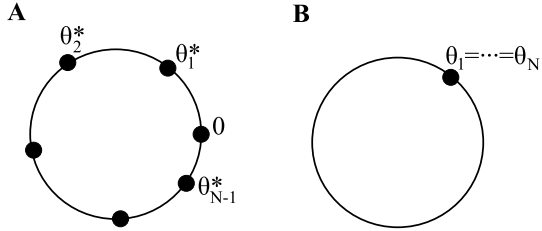}
	\caption{Two features of kick synchronization. A. The oscillators asymptotically converge to a unique phase-locked configuration. B. The oscillators achieve full synchronization in finite time.}
	\label{phase_lock_synchro}
\end{figure}

\begin{enumerate}[(i)]
	\item \emph{Isolated phase-locked configuration:} If the phase response $Z_\epsilon$ is monotone decreasing, the network globally converges toward the unique fixed point of the $(N-1)$-dimensional firing map, which corresponds to the \emph{unique phase-locked configuration} of $N$ oscillators (\myfigurename~\ref{phase_lock_synchro}A). This behavior was previously obtained in \myfigurename~\ref{impulsive_vdp_phaselocked} with two van der Pol oscillators in the relaxation limit. Except in the weak coupling limit, the phase-locked configuration is not a splay state (that is, the phase differences between the successive oscillators are not identical). In addition, since a single oscillator may represent a cluster of (locally synchronized) oscillators, this configuration also corresponds to a phase-locked clustering configuration.

	\item \emph{Finite-time synchronization:} If the phase response~$Z_\epsilon$ is monotone increasing, Theorem~\ref{theo_fir_map} implies that the fixed point is (globally) unstable. Then, successive absorptions lead the network to \emph{full synchronization in finite time}, that is, all oscillators share the same phase (\myfigurename~\ref{phase_lock_synchro}B) (see also~\cite{Mirollo:1990ft}). This behavior was previously obtained in \myfigurename~\ref{impulsive_vdp_synchro} with two van der Pol oscillators in the relaxation limit.
\end{enumerate}

% \paragraph*{(i) Isolated phase-locked configuration} If the phase response $Z_\epsilon$ is monotone decreasing, the network globally converges toward the unique fixed point of the $(N-1)$-dimensional firing map, which corresponds to the \emph{unique phase-locked configuration} of $N$ oscillators (\myfigurename~\ref{phase_lock_synchro}A). This behavior was previously obtained in \myfigurename~\ref{impulsive_vdp_phaselocked} with two van der Pol oscillators in the relaxation limit. Except in the weak coupling limit, the phase-locked configuration is not a splay state (that is, the phase differences between the successive oscillators are not identical). In addition, since a single oscillator may represent a cluster of (locally synchronized) oscillators, this configuration also corresponds to a phase-locked clustering configuration.
% 
% \paragraph*{(ii) Finite-time synchronization} If the phase response~$Z_\epsilon$ is monotone increasing, Theorem~\ref{theo_fir_map} implies that the fixed point is (globally) unstable. Then, successive absorptions lead the network to \emph{full synchronization in finite time}, that is, all oscillators share the same phase (\myfigurename~\ref{phase_lock_synchro}B) (see also~\cite{Mirollo:1990ft}). This behavior was previously obtained in \myfigurename~\ref{impulsive_vdp_synchro} with two van der Pol oscillators in the relaxation limit.

The study of kick synchronization for general phase dynamics is still (very) limited. To our knowledge, there is so far no global result for oscillators that are not characterized by a monotone PRC. While some oscillators satisfy the hypotheses of Theorem~\ref{theo_fir_map} (\eg{}~LIF oscillators, van der Pol oscillators in the relaxation limit $\mu\gg1$), other oscillators are not characterized by a monotone PRC (\eg{}~QIF oscillators). In this latter case, even a local stability analysis may become elusive---although Assumption~\ref{order_preserv} ensures that the firing map has still a unique fixed point---and the networks can display other (more complex) collective behaviors~\cite{Mauroy:2010wca}. In addition, the general study of oscillators that do not satisfy Assumption~\ref{order_preserv}---for which a firing map cannot be defined---remains an open problem. These few examples are all relevant research perspectives.

Whereas few studies have investigated kick synchronization for oscillators with general phase dynamics, several extensions of the original impulsive coupling can be found in literature (\eg{}~reduced interconnectivity~\cite{Denker:2004vc,DiazGuilera:1997wt,Dror:1999gn,Goel:2002ew,Ostborn:2002wh,Timme:2002vu,Timme:2004tx}, delays~\cite{Ernst:1998uy,Timme:2002uk}, non-instantaneous interactions~\cite{Abbott:1993vb,Bressloff:2000wn,vanVreeswijk:1994wy,Zillmer:2007kl}, non-identical oscillators~\cite{Bottani:1996tz,Chang:2008de,DeSmet:2010wt,Senn:2000tx}).

\subsection{Infinite populations}

Kick synchronization can also be studied in the continuous limit of $N \rightarrow \infty$ oscillators. In this case, a continuum of oscillators is described by a phase density function $\rho(\theta,t)$ normalized on~$\mathbb{S}^1$, or is equivalently described by a flux
\begin{equation*}
	J(\theta,t)=\rho(\theta,t)\, v(\theta,t)\,,
\end{equation*}
where $v(\theta,t)$ is the velocity of the oscillators. The evolution of the oscillators obeys the continuity equation
\begin{equation}
	\label{cont_eq}
	\frac{\partial}{\partial t}\rho(\theta,t)=-\frac{\partial}{\partial \theta} J(\theta,t)\,,
\end{equation}
with the boundary condition $J(0,t)=J(2\pi,t)$ $\forall t$.

When the population is infinite, the impulsive coupling is continuous and proportional to the flux $J(2\pi,t)$. We derive the result as follows. In the case of finite populations, the flow rule~\eqref{flow_rule} and the jump rule~\eqref{jump_rule} imply a velocity
\begin{equation}
\label{2rules_velocity}
	v(\theta_i,t)=\omega+Z_{\epsilon}(\theta_i)\,\sum_{j \in \mathcal{N}\setminus\{i\}} \sum_{k=0}^{\infty} \delta(t-t^{(j)}_k)\,,
\end{equation}
where $t^{(j)}_k$ denote the times at which oscillator $j$ reaches the phase $\theta=2\pi$. Since the flux is given by $J(2\pi,t)=1/N \sum_{j\in \mathcal{N}} \sum_{k=0}^{\infty} \delta(t-t^{(j)}_k)$, it follows that, for a large number of oscillators where $\mathcal{N}\setminus\{i\}\approx \mathcal{N}$, one has
\begin{equation}
	\label{velocity_finite}
	v(\theta_i,t) \approx \omega + Z_{\epsilon}(\theta_i)\,N J(2\pi,t)\,.
\end{equation}
In the limit of an infinite number of oscillators, the impulsive coupling is an infinite sum of infinitesimal kicks $\epsilon=K/N \ll 1$, where $K$ is a positive constant. Then, Remark~\ref{PRC_weak_coupling} and~\eqref{velocity_finite} imply that
\begin{equation}
	\label{continuous_velocity}
	v(\theta,t)=\omega+K\,Z(\theta)\,J(2\pi,t)
\end{equation}
and the coupling is proportional to the (continuous) flux~$J(2\pi,t)$.

%The impulsive coupling is continuous: defined as an infinite sum of infinitesimal kick $\epsilon=K/N$, where $K$ is a positive constant, the impulsive coupling is proportional to the flux at phase $\theta=2\pi$. The velocity of the oscillators is given by
%\begin{equation}
%\label{continuous_velocity}
%v(\theta,t)=\omega+K\,Z(\theta)\,J(2\pi,t)
%\end{equation}
%and is the continuous equivalent of~\eqref{flow_rule} and~\eqref{jump_rule}. It is noticeable that the iPRC $Z$ is used instead of $Z_\epsilon$, since $\epsilon=K/N\rightarrow 0$ ??(see eq ).??

As a parallel to the results obtained for finite populations, the continuity equation~\eqref{cont_eq}--\eqref{continuous_velocity} has strong stability properties if the iPRC is monotone. Its global stability is shown using the continuous analog of the $1$-norm~\eqref{1_norm}, which has the interpretation of a total variation distance. The result is summarized as follows~\cite{Mauroy:2011ux}:
\begin{theorem}
	\label{theo_EDP}
	Consider an iPRC that satisfies $Z''(\theta)>0$ $\forall \theta\in(0,2\pi)$ or $Z''(\theta)<0$ $\forall \theta\in(0,2\pi)$. Then, an admissible solution of~\eqref{cont_eq}--\eqref{continuous_velocity}
	\begin{itemize}
		\item exponentially converges to the unique stationary solution (when it exists) if $Z'(\theta)<0$ $\forall \theta\in(0,2\pi)$;
		\item reaches synchronization (infinite flux) in finite time if $Z'(\theta)>0$ $\forall \theta\in(0,2\pi)$. \hfill $\lrcorner$
	\end{itemize}
\end{theorem}

\begin{figure}
	\centering
	\subfigure{\includegraphics[width=4.5cm]{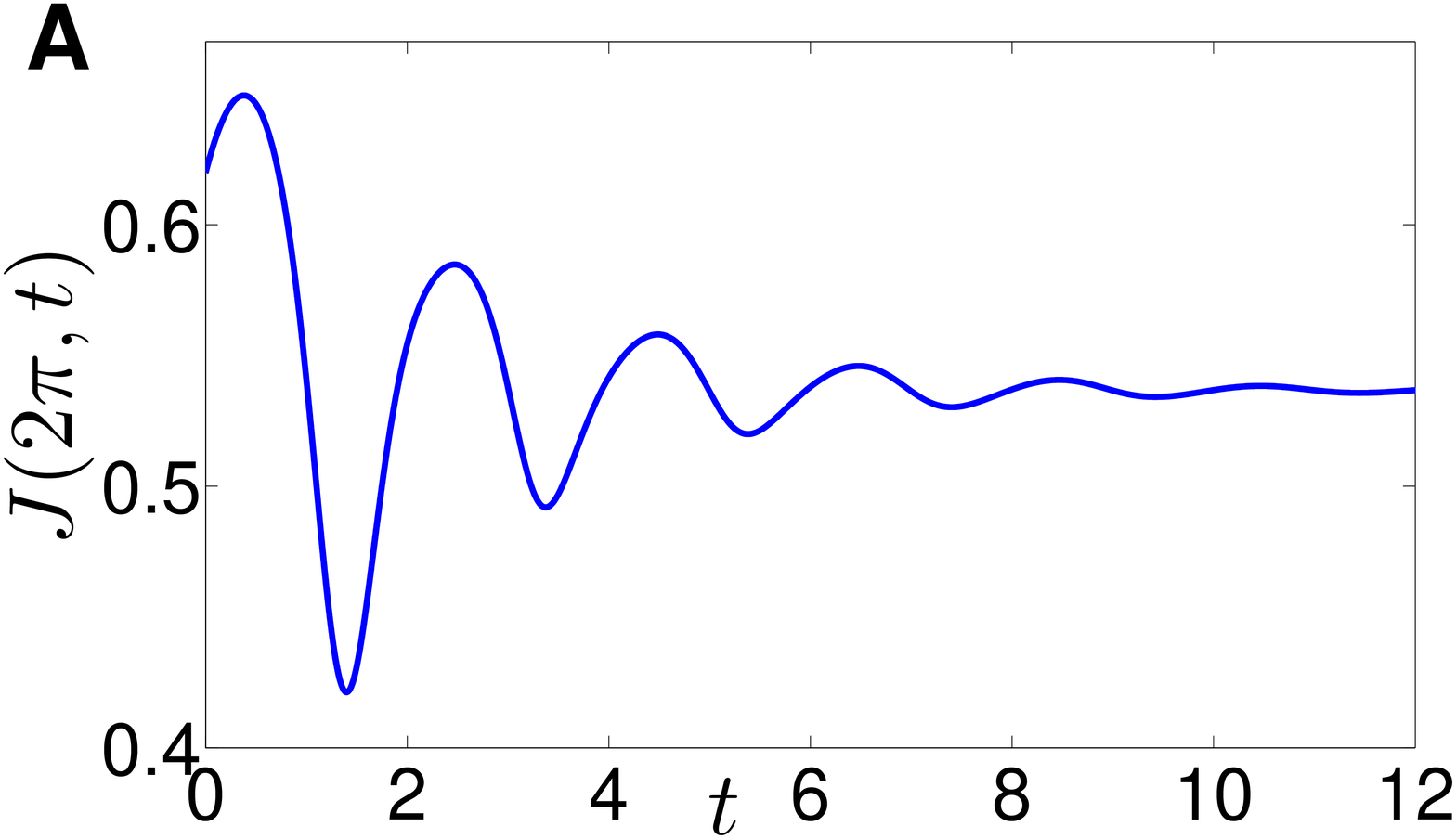}}
	\subfigure{\includegraphics[width=4.5cm]{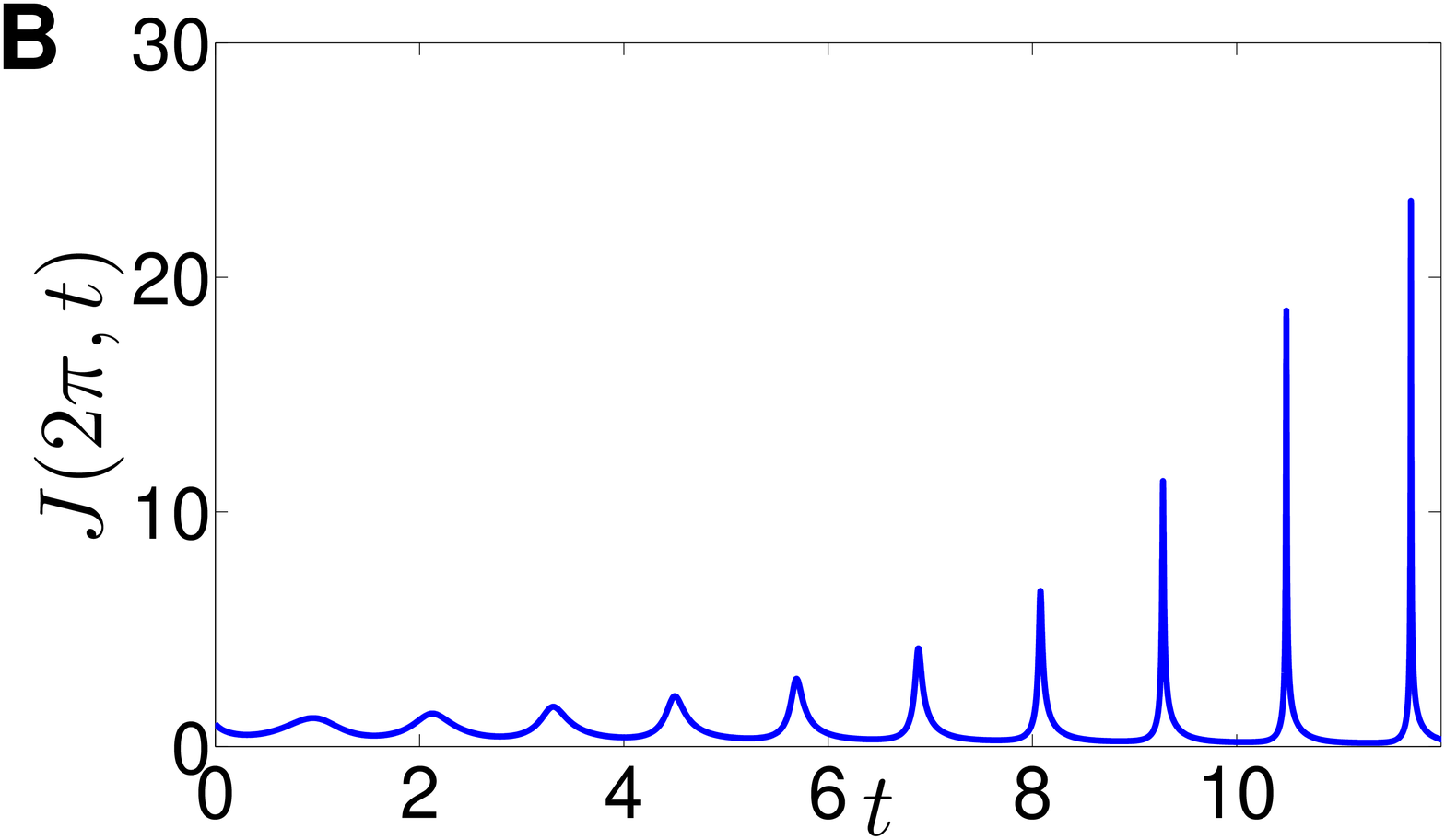}}
	\caption{Two collective behaviors for kick synchronization in infinite populations. A. The continuum converges to the stationary solution characterized by $J(\theta,t)=J^*$. B. The continuum reaches synchronization in finite time.}
	\label{continu}
\end{figure}

According to Theorem~\ref{theo_EDP}, kick synchronization in infinite populations involves two collective behaviors, which are the exact analogs of the behaviors observed for finite populations. If the iPRC is monotone decreasing, the network converges to the \emph{unique stationary solution}, which corresponds to a constant flux $J(\theta)=J^*$ (\myfigurename~\ref{continu}A). This is the analog of the phase-locked clustering configuration. If the iPRC is monotone increasing, the flux tends to a Dirac function and the network achieves \emph{synchronization in finite time} (\myfigurename~\ref{continu}B).

LIF oscillators and van der Pol oscillators in the relaxation limit $\mu \gg 1$ are characterized by a monotone iPRC that satisfies the hypotheses of Theorem~\ref{theo_EDP}. If the iPRC is not monotone, there is so far no global stability result for the continuity equation~\eqref{cont_eq}--\eqref{continuous_velocity}. However, local stability of the stationary solution can be studied numerically (see \eg{}~\cite{vanVreeswijk:1996wb}).

%%%%%%% SECT.4 %%%%%%%
\section{Phase models in the weak coupling limit} \label{sec:weak}

The results presented in the previous sections emphasize  important differences between diffusive synchronization and impulsive synchronization. Both the analysis techniques and the contraction measures differ in a fundamental way. However, it is remarkable that both frameworks are unified in the weak coupling limit. Indeed, using averaging techniques, Kuramoto showed that the phase dynamics of weakly coupled oscillators can always be reduced to the unique canonical form
\begin{equation}
	\label{gen_form_Kuramoto}
	\dot{\theta}_i=\omega+\sum_{j\in \mathcal{N}_i} \couplfun_{ij}(\theta_i-\theta_j) \,,
\end{equation}
where the $\couplfun_{ij}(\cdot)$ are coupling functions closely related to the iPRC~\cite{Kuramoto:1984wo}. As a consequence, the phase dynamics~\eqref{gen_form_Kuramoto} are a general paradigm that holds both for diffusively coupled oscillators and impulsively coupled oscillators.

\subsection{Weak diffusive coupling}
\label{subsec_weak_diffusive}

In the case of a weak diffusive coupling, we show that the phase dynamics~\eqref{gen_form_Kuramoto} correspond to the averaged dynamics of the oscillators. (We follow similar lines as in~\cite{Kuramoto:1984wo} (see also~\cite{Hoppensteadt:1997tp})).

It follows from~\eqref{eq:weak_coupling} and~\eqref{eq:diffusive_coupling} that the oscillators are characterized by the phase dynamics
\begin{equation*}
	\dot{\theta}_i=\omega+\epsilon Z(\theta_i) \sum_{j\in\mathcal{N}_i} K_{ji} (\tilde{\vecH}_j(\theta_j)-\tilde{\vecH}_i(\theta_i))\,,
\end{equation*}
with $\tilde{\vecH}_i(\theta_i)=\vecH_i(\vecx^\gamma(\theta_i/\omega))$. The phases can be decomposed as $\theta_i=\omega t+\psi_i$, where $\psi_i$ are slow phase deviations from the uniform natural oscillation $\omega\,t$. Then, the phase dynamics are rewritten as
\begin{equation*}
	\dot{\psi}_i=\epsilon Z(\omega t+\psi_i) \sum_{j \in \mathcal{N}_i} K_{ji} (\tilde{\vecH}_j(\omega t+\psi_j)-\tilde{\vecH}_i(\omega t+\psi_i))\,.
\end{equation*}
Note that $Z$ is considered here as the $2\pi$-periodic extension of the iPRC on the real line, \ie{}~$Z(x)\equiv Z(x \bmod 2\pi)$. Next, averaging the above dynamics over a period $T=2\pi/\omega$ and under fixed $\psi_i$ and $\psi_j$, we obtain
\begin{align*}
		\dot{\psi}_i & = \sum_{j \in \mathcal{N}_i} \frac{\epsilon K_{ji}}{T} \int_0^T Z(\omega t+\psi_i) \times (\tilde{\vecH}_j(\omega t+\psi_j)-\tilde{\vecH}_i(\omega t+\psi_i)) \, dt \\
		& = \sum_{j \in \mathcal{N}_i} \frac{\epsilon K_{ji}}{\omega} \int_0^{2\pi} Z(\psi_i-\psi_j+s) \times (\tilde{\vecH}_j(s)-\tilde{\vecH}_i(\psi_i-\psi_j+s)) \, ds
\end{align*}
where we have used the change of variable $\omega t + \psi_j=s$. With the coupling functions
\begin{equation} \label{eq:couplfun-diff}
	\couplfun_{ij}(\cdot)= \frac{\epsilon K_{ji}}{\omega} \int_0^{2\pi} Z(\cdot+s) \, (\tilde{\vecH}_j(s)-\tilde{\vecH}_i(\cdot+s)) \, ds \,,
\end{equation}
the last equation yields
\begin{equation}
	\label{perturb_gen_form}
	\dot{\psi}_i=\sum_{j\in \mathcal{N}_i} \couplfun_{ij}(\psi_i-\psi_j)\,,
\end{equation}
which is equivalent to~\eqref{gen_form_Kuramoto}.

\begin{remark}[van der Pol oscillator]
	Mimicking the computation steps to transform \eqref{eq:vanderpol} to \eqref{PRC_1}, we can write \eqref{eq:vdp-diff} in polar coordinates $(x,w) = (r\sin(\phi),-r\cos(\phi))$ and apply averaging theory to obtain the phase dynamics given by
	\begin{equation*}
		\dot{\theta} = \omega + \epsilon \underbrace{\frac{\omega \cos(\phi)}{\tilde{r}+\mu g(\tilde{r}\sin(\phi))\cos(\phi)}}_{\reveqdef Z(\theta)} \bar{u}
	\end{equation*}
	with $\phi$ given by the appropriate bijective change of variable $\Theta^{-1}(\theta)$, $\tilde{r} = 2 + \mathcal{O}(\mu,\epsilon)$, and $g(x) = (x-x^3/3)$.
	
	For values of $\mu$ tending to 0, the oscillator output is given by $\tilde{H}(\theta) = 2\sin(\theta)$ and the iPRC by $Z(\theta) = \frac{1}{2}\cos(\theta)$. Applying \eqref{eq:couplfun-diff} and using basic trigonometry, the coupling function is given by $\couplfun_{ij}(\theta) = -\epsilon K_{ji}\pi\sin(\theta)$. 
	The diffusive interconnection of quasi-harmonic van der Pol oscillators leads thus naturally to the popular Kuramoto model characterized by a sinusoidal coupling function.
	\hfill $\lrcorner$
\end{remark}

\subsection{Weak impulsive coupling}

Next, we derive the averaged phase dynamics~\eqref{gen_form_Kuramoto} in the case of a weak impulsive coupling (see also~\cite{Kuramoto:1991ja}). From~\eqref{PRC_weak_coupling} and~\eqref{2rules_velocity}, one has
\begin{equation*}
	\dot{\theta_i}=\omega+\epsilon Z(\theta_i)\,\sum_{j \in \mathcal{N}_i} K_{ji} \sum_{k=0}^{\infty} \delta(t-t^{(j)}_k)\,,
\end{equation*}
where, in full generality, we have introduced the constants $K_{ji}$ and a general interconnection topology $\mathcal{N}_i$. As in Section~\ref{subsec_weak_diffusive}, using the phase deviation $\psi_i$ and averaging the dynamics under a constant $\psi_i$ yield
\begin{equation*}
	\begin{split}
		\dot{\psi_i}& =\sum_{j \in \mathcal{N}_i} \frac{\epsilon K_{ji}}{T} \int_0^T Z(\omega t+\psi_i) \sum_{k=0}^{\infty} \delta(t-t^{(j)}_k)\, dt  \\
		& = \sum_{j \in \mathcal{N}_i} \frac{\epsilon K_{ji}}{T} Z(\omega t^{(j)}+\psi_i) \\
		& = \sum_{j \in \mathcal{N}_i} \frac{\epsilon K_{ji}}{T} Z(\psi_i-\psi_j)\,,
	\end{split}
\end{equation*}
where the last equality holds since $\omega t^{(j)} + \psi_j=2\pi$ and given the periodicity of the iPRC. With the coupling functions
\begin{equation}
	\label{coupling_fct_impulsive}
	\couplfun_{ij}(\cdot)=\frac{\epsilon K_{ji}}{T} Z(\cdot)\,,
\end{equation}
we recover~\eqref{perturb_gen_form}, which corresponds to the general form~\eqref{gen_form_Kuramoto}. In the case of a (weak) impulsive coupling, it is remarkable that the coupling functions are proportional to the iPRC itself.

\subsection{Global stability results and collective behaviors}

In the weak coupling limit, the collective behaviors of diffusively coupled and impulsively coupled oscillators depend on the (global) stability properties of the averaged dynamics~\eqref{gen_form_Kuramoto}. When the oscillators are initialized within a semicircle, a change of coordinate maps the dynamics into~$\mathbb{R}^N$ and the analysis of~\eqref{gen_form_Kuramoto} is equivalent to a consensus problem on a convex set (see \eg{}~\cite{Moreau:2005km}). In this case, a global stability analysis can be performed using consensus theory~\cite{Moreau:2005km}. In this section, we rather discuss the global stability properties of~\eqref{gen_form_Kuramoto} on the whole (nonconvex) torus~$\mathbb{T}^N$, but for the particular all-to-all topology with identical connections (\ie{}~$\mathcal{N}_i=\mathcal{N}\setminus\{i\}$ and $K_{ji}=K$ $\forall i,j$).

\subsubsection{Kuramoto model} The most popular model of the form~\eqref{gen_form_Kuramoto} is the Kuramoto model~\cite{Kuramoto:1984dl}. Characterized by the coupling function $\couplfun(\theta)=-K/N \sin(\theta)$, which can be considered as the first Fourier harmonic of a more complex coupling function, the Kuramoto model appears as the generic model for the averaged dynamics of (diffusively) coupled oscillators. 
% \alert{In particular, it corresponds to the phase model of van der Pol oscillators in the quasi-harmonic limit with a weak diffusive coupling (see ??).}

The Kuramoto model is characterized by the following collective behaviors (for almost every initial condition):
\begin{itemize}
	\item If $K<0$ (inhibitory coupling), the oscillators converge toward an incoherent state characterized by $\sum_{k\in\mathcal{N}} e^{i\theta_k}=0$ (with $i=\sqrt{-1}$) (balanced state);
	\item If $K>0$ (excitatory coupling), the oscillators achieve full synchronization.
\end{itemize}
These global properties result from the fact that~\eqref{gen_form_Kuramoto} is a gradient system for Kuramoto model~\cite{Hoppensteadt:1997tp}. In the case of shifted coupling functions $\couplfun(\theta)=\sin(\theta-\delta)$, a general Lyapunov function still enforces global convergence toward synchronization or incoherent state~\cite{Watanabe:1994hh}.

\subsubsection{Monotone coupling function} It follows from~\eqref{coupling_fct_impulsive} that a monotone coupling function is obtained in the case of an impulsive coupling, for oscillators characterized by a monotone iPRC (\eg{}~LIF oscillators, van der Pol oscillators in the relaxation limit $\mu\gg 1$). (The coupling function is monotone on $(0,2\pi)$ and characterized by a discontinuity $\couplfun(0^-) \neq \couplfun(0^+)$, since it satisfies the $2\pi$-periodicity condition.) Then, as was shown in Section~\ref{sec:kick}, the monotonicity property of the oscillators induces a contraction property for the averaged dynamics~\eqref{gen_form_Kuramoto}.

The contraction is shown in a rotating frame associated with an oscillator, that is, for the equivalent dynamics
\begin{equation}
	\label{phase_dyn_modified}
	\dot{\varphi}_i=\couplfun(\varphi_i)+\sum_{\substack{j=1\\j\neq i}}^{N-1} \couplfun(\varphi_i-\varphi_j)- \sum_{j=1}^{N-1} \couplfun(-\varphi_j)
\end{equation}
that are obtained by using the change of variable $\varphi_i=\theta_{i+1}-\theta_1$ in~\eqref{gen_form_Kuramoto}. Note that~\eqref{phase_dyn_modified} is defined in the closure of the cone $\mathcal{C}=\{(\varphi_1,\cdots,\varphi_{N-1}) \in (0,2\pi)^{N-1}|\varphi_i < \varphi_{i+1} \}$. The result, which is the analog of Theorem~\ref{theo_fir_map}, is summarized in the following theorem~\cite{Mauroy:2012ts}.
\begin{theorem}
	\label{theo_contract_monot}
	Consider a monotone coupling function that satisfies $\couplfun''(\theta)>0$ $\forall \theta\in(0,2\pi)$ or $\couplfun''(\theta)<0$ $\forall \theta\in(0,2\pi)$. Then,
	\begin{itemize}
		\item the dynamics~\eqref{phase_dyn_modified} are contracting in $\mathcal{C}$ with respect to~\eqref{1_norm} if $\couplfun'(\theta)<0$ $\forall \theta\in(0,2\pi)$;
		\item the dynamics~\eqref{phase_dyn_modified} are expanding in $\mathcal{C}$ with respect to~\eqref{1_norm} if $\couplfun'(\theta)>0$ $\forall \theta\in(0,2\pi)$. \hfill $\lrcorner$
	\end{itemize}
\end{theorem}

A corollary of the contraction property of Theorem~\ref{theo_contract_monot} is that networks of oscillators coupled through a monotone coupling function display two behaviors (for almost every initial condition):
\begin{itemize}
	\item If $\couplfun$ is monotone decreasing, the oscillators converge to the \emph{unique incoherent configuration} $\varphi^*_k=k\frac{2\pi}{N}$ (splay state);
	\item If $\couplfun$ is monotone increasing, the oscillators achieve \emph{full synchronization in finite time}.
\end{itemize}

This global behavior is similar to the global behavior of Kuramoto oscillators, but also characteristic of kick synchronization. The synchronization takes place in finite time, and the incoherent state is an isolated fixed point (splay state), whereas it is a $(N-3)$-dimensional manifold in Kuramoto model. A consequence of this difference is that the asymptotic behavior of~\eqref{gen_form_Kuramoto} is robust to small heterogeneity in the natural frequencies $\omega$ in monotone firing oscillators~\cite{Mauroy:2008dp} whereas the asymptotic dynamics of Kuramoto model can be highly complex even for small heterogeneities~\cite{Strogatz:2000wx}.

\subsubsection{Other coupling function} As for pulse-coupled models, there are only few global stability results for phase models characterized by generic---non-sinusoidal, non-monotone---coupling functions. When the coupling function is odd, the phase oscillators correspond to a gradient system and are characterized by global properties similar to Kuramoto model~\cite{Hoppensteadt:1997tp}. For more general coupling functions, however, a global stability analysis is usually elusive (\eg{}~weakly pulse-coupled QIF oscillators~\cite{Mauroy:2011th}). In this case, a local stability analysis can be performed and, in particular, there exists a criterion for the local stability of the incoherent state~\cite{Brown:2003vn,Kuramoto:1991ja}. In this context, local stability of clustering configurations has been investigated in~\cite{Golomb:1992ia,Okuda:1993dj}, for instance, and the design of the coupling function to achieve particular cluster states is considered in~\cite{Orosz:2009wx}. Note also that several studies have considered generic coupling functions in the case of non-identical phase oscillators (\eg{}~\cite{Daido:1992jx}).

%%%%%%% SECT.5 %%%%%%%

%\section{Contraction and synchronization}
%
%Slotine paper. 
%Sync by entrainment. Russo/Sontag.
%QIF conjecture.

%%% Sec 5.A

%%% Sec 5.C

%%%% Sec 5.E
 
%%%%%%% CONCLUSION %%%%%%%
\section{Conclusion} \label{sec:conclusion}

Diffusive synchronization and kick synchronization are two distinctively different models of synchronization, underlying different physical synchronization mechanisms and leading to different analysis tools. While most popular manifestations of synchronization seem more akin to the kick model than to the diffusive model, the literature on kick synchronization is sparse, probably owing to the hybrid nature of the kick model and to the mathematical difficulty of analyzing nonlinear resonance. For instance, a number of kick synchronization problems remain unsolved (\eg{}~non-identical oscillators) and several research themes are still unexplored (\eg{}~general---application oriented---dynamics).

By essence, both the diffusive model and the kick model are crude idealizations of the complex synchronization phenomena observed in nature. But it is fair to recognize that the importance of the `kick' deserves more consideration in a mathematical literature dominated by the diffusive model. This is for instance illustrated in the recent paper celebrating the 300th anniversary of the first scientific investigation of synchronization by Huygens~\cite{Bennett:2002uo}. In that sense, the present paper is an invitation to the growing hybrid systems community to contribute---as in the recent work~\cite{Nunez:2012um}---to a deeper understanding of a fundamental property of interconnected nonlinear dynamical systems.

%%%%%%% THANKS %%%%%%%

\section{Acknowledgments}

This paper presents research results of the Belgian Network DYSCO (Dynamical Systems, Control, and Optimization), funded by the Interuniversity Attraction Poles Programme, initiated by the Belgian State, Science Policy Office. The scientific responsibility rests with its authors. A.~Mauroy holds a postdoctoral fellowship from the Belgian American Educational Foundation. P.~Sacr\'{e} is supported as an F.R.S.-FNRS Research Fellow (Belgian Fund for Scientific Research).

%%%%%%% BIBLI %%%%%%%

\bibliographystyle{abbrvurl}
\bibliography{./bib/ps-bibfile}

\begin{thebibliography}{10}

\bibitem{Abbott:1999ui}
L.~F. Abbott.
\newblock {Lapicque's introduction of the integrate-and-fire model neuron
  (1907)}.
\newblock {\em Brain Res. Bull.}, 50(5-6):303--304, Nov. 1999.

\bibitem{Abbott:1993vb}
L.~F. Abbott and C.~van Vreeswijk.
\newblock {Asynchronous states in networks of pulse-coupled oscillators}.
\newblock {\em Phys. Rev. E}, 48(2):1483--1490, Aug. 1993.

\bibitem{Angeli:2002eb}
D.~Angeli.
\newblock {A Lyapunov approach to incremental stability properties}.
\newblock {\em IEEE Trans. Autom. Control}, 47(3):410--421, Mar. 2002.

\bibitem{Arcak:2007dq}
M.~Arcak.
\newblock {Passivity as a design tool for group coordination}.
\newblock {\em IEEE Trans. Autom. Control}, 52(8):1380--1390, Aug. 2007.

\bibitem{Bennett:2002uo}
M.~R. Bennett, M.~F. Schatz, H.~Rockwood, and K.~Wiesenfeld.
\newblock {Huygens's clocks}.
\newblock {\em Proc. R. Soc. Lond. A}, 458:563--579, Mar. 2002.

\bibitem{Blekhman:1988wn}
I.~I. Blekhman.
\newblock {\em {Synchronization in Science and Technology}}.
\newblock American Society of Mechanical Engineers, 1988.

\bibitem{Bottani:1996tz}
S.~Bottani.
\newblock {Synchronization of integrate and fire oscillators with global
  coupling}.
\newblock {\em Phys. Rev. E}, 54(3):2334--2350, Sept. 1996.

\bibitem{Bressloff:2000wn}
P.~C. Bressloff and S.~Coombes.
\newblock {A dynamical theory of spike train transitions in networks of
  integrate-and-fire oscillators}.
\newblock {\em SIAM J. Appl. Math.}, 60(3):820--841, 2000.

\bibitem{Brown:2003vn}
E.~T. Brown, P.~Holmes, and J.~Moehlis.
\newblock {Globally coupled oscillator networks}.
\newblock In E.~Kaplan, J.~E. Marsden, and K.~R. Sreenivasan, editors, {\em
  Perspectives and Problems in Nonlinear Science: a Celebratory Volume in Honor
  of Larry Sirovich}, pages 183--215. Springer, New York, NY, 2003.

\bibitem{Brown:2004iy}
E.~T. Brown, J.~Moehlis, and P.~Holmes.
\newblock {On the phase reduction and response dynamics of neural oscillator
  populations}.
\newblock {\em Neural Comput.}, 16(4):673--715, Apr. 2004.

\bibitem{Buck:1988ui}
J.~B. Buck.
\newblock {Synchronous rhythmic flashing of fireflies. II.}
\newblock {\em Q. Rev. Biol.}, 63(3):265--289, Sept. 1988.

\bibitem{Chang:2008de}
Y.-C. Chang and J.~Juang.
\newblock {Stable synchrony in globally coupled integrate-and-fire
  oscillators}.
\newblock {\em SIAM J. Appl. Dyn. Syst.}, 7(4):1445--1476, 2008.

\bibitem{Chopra:2012wo}
N.~Chopra.
\newblock {Output synchronization on strongly connected graphs}.
\newblock {\em IEEE Trans. Autom. Control}, PP(99):1--1, Apr. 2012.

\bibitem{Daido:1992jx}
H.~Daido.
\newblock {Order function and macroscopic mutual entrainment in uniformly
  coupled limit-cycle oscillators}.
\newblock {\em Prog. Theor. Phys.}, 88(6):1213--1218, Dec. 1992.

\bibitem{DeSmet:2010wt}
F.~De~Smet and D.~Aeyels.
\newblock {Coexistence of stable stationary behavior and partial synchrony in
  an all-to-all coupled spiking neural network}.
\newblock {\em Phys. Rev. E}, 82(6 Pt 2):066208, Dec. 2010.

\bibitem{Denker:2004vc}
M.~Denker, M.~Timme, M.~Diesmann, F.~Wolf, and T.~Geisel.
\newblock {Breaking synchrony by heterogeneity in complex networks}.
\newblock {\em Phys. Rev. Lett.}, 92(7):074103, Feb. 2004.

\bibitem{DiazGuilera:1997wt}
A.~D{\'\i}az-Guilera, A.~Arenas, A.~Corral, and C.~J. P{\'e}rez.
\newblock {Stability of spatio-temporal structures in a lattice model of
  pulse-coupled oscillators}.
\newblock {\em Physica D}, 103(1-4):419--429, Apr. 1997.

\bibitem{Dror:1999gn}
R.~O. Dror, C.~C. Canavier, R.~J. Butera, J.~W. Clark, and J.~H. Byrne.
\newblock {A mathematical criterion based on phase response curves for
  stability in a ring of coupled oscillators}.
\newblock {\em Biol. Cybern.}, 80(1):11--23, 1999.

\bibitem{Ermentrout:1986ve}
G.~B. Ermentrout and N.~Kopell.
\newblock {Parabolic bursting in an excitable system coupled with a slow
  oscillation}.
\newblock {\em SIAM J. Appl. Math.}, 46(2):233--253, 1986.

\bibitem{Ernst:1998uy}
U.~Ernst, K.~Pawelzik, and T.~Geisel.
\newblock {Delay-induced multistable synchronization of biological
  oscillators}.
\newblock {\em Phys. Rev. E}, 57(2):2150--2162, 1998.

\bibitem{Farkas:1994uq}
M.~Farkas.
\newblock {\em {Periodic Motions}}, volume 104 of {\em Applied Mathematical
  Sciences}.
\newblock Springer-Verlag, New York, NY, 1994.

\bibitem{Fitzhugh:1961il}
R.~Fitzhugh.
\newblock {Impulses and physiological states in theoretical models of nerve
  membrane}.
\newblock {\em Biophys. J.}, 1(6):445--466, July 1961.

\bibitem{Gerstner:2002ti}
W.~Gerstner and W.~M. Kistler.
\newblock {\em {Spiking Neuron Models: Single Neurons, Populations,
  Plasticity}}.
\newblock Cambridge University Press, Cambridge, England, Aug. 2002.

\bibitem{Glass:1988ub}
L.~Glass and M.~C. Mackey.
\newblock {\em {From Clocks to Chaos: the Rhythms of Life}}.
\newblock Princeton University Press, Princeton, NJ, 1988.

\bibitem{Goel:2002ew}
P.~Goel and G.~B. Ermentrout.
\newblock {Synchrony, stability, and firing patterns in pulse-coupled
  oscillators}.
\newblock {\em Physica D}, 163(3-4):191--216, 2002.

\bibitem{Goldbeter:1996uo}
A.~Goldbeter.
\newblock {\em {Biochemical Oscillations and Cellular Rhythms: the Molecular
  Bases of Periodic and Chaotic Behaviour}}.
\newblock Cambridge University Press, Cambridge, United Kingdom, 1996.

\bibitem{Golomb:1992ia}
D.~Golomb, D.~Hansel, B.~Shraiman, and H.~Sompolinsky.
\newblock {Clustering in globally coupled phase oscillators}.
\newblock {\em Phys. Rev. A}, 45(6):3516--3530, Mar. 1992.

\bibitem{Guckenheimer:1983up}
J.~Guckenheimer and P.~Holmes.
\newblock {\em {Nonlinear Oscillations, Dynamical Systems, and Bifurcations of
  Vector Fields}}, volume~42 of {\em Applied Mathematical Sciences}.
\newblock Springer, New York, NY, 2nd edition, 1983.

\bibitem{Hamadeh:2011eu}
A.~Hamadeh, G.-B. Stan, R.~Sepulchre, and J.~Gon{\c c}alves.
\newblock {Global state synchronization in networks of cyclic feedback
  systems}.
\newblock {\em IEEE Trans. Autom. Control}, 57(2):478--483, Feb. 2012.

\bibitem{Hodgkin:1952td}
A.~L. Hodgkin and A.~F. Huxley.
\newblock {A quantitative description of membrane current and its application
  to conduction and excitation in nerve}.
\newblock {\em J. Physiol.}, 117(4):500--544, Aug. 1952.

\bibitem{Hong:2005jw}
Y.-W. Hong and A.~Scaglione.
\newblock {A scalable synchronization protocol for large scale sensor networks
  and its applications}.
\newblock {\em IEEE J. Sel. Areas Commun.}, 23(5):1085--1099, 2005.

\bibitem{Hoppensteadt:1997tp}
F.~C. Hoppensteadt and E.~M. Izhikevich.
\newblock {\em {Weakly Connected Neural Networks}}, volume 126 of {\em Applied
  Mathematical Sciences}.
\newblock Springer-Verlag, New York, NY, 1997.

\bibitem{Huygens:1673tm}
C.~Huygens.
\newblock {\em {Horologium Oscillatorium}}.
\newblock Apud F. Muguet, Parisiis, France, 1673.

\bibitem{Huygens:1673vi}
C.~Huygens.
\newblock {\em {Oeuvres Compl{\`e}tes de Christiaan Huygens}}, volume~17.
\newblock Martinus Nijhoff, The Hague, The Netherlands, 1932.

\bibitem{Izhikevich:2007vr}
E.~M. Izhikevich.
\newblock {\em {Dynamical Systems in Neuroscience: the Geometry of Excitability
  and Bursting}}.
\newblock The MIT Press, Cambridge, MA, 2007.

\bibitem{Khalil:2002wj}
H.~K. Khalil.
\newblock {\em {Nonlinear Systems}}.
\newblock Prentice Hall, Upper Saddle River, NJ, 3rd edition, 2002.

\bibitem{Knight:1972hu}
B.~W. Knight.
\newblock {Dynamics of encoding in a population of neurons}.
\newblock {\em The Journal of General Physiology}, 59(6):734--766, June 1972.

\bibitem{Kuramoto:1975ki}
Y.~Kuramoto.
\newblock {Self-entrainment of a population of coupled non-linear oscillators}.
\newblock In {\em International Symposium on Mathematical Problems in
  Theoretical Physics}, pages 420--422. Springer-Verlag, Berlin/Heidelberg,
  Germany, 1975.

\bibitem{Kuramoto:1984wo}
Y.~Kuramoto.
\newblock {\em {Chemical Oscillations, Waves, and Turbulence}}, volume~19 of
  {\em Springer Series in Synergetics}.
\newblock Springer-Verlag, Berlin/Heidelberg, Germany, 1st edition, 1984.

\bibitem{Kuramoto:1984dl}
Y.~Kuramoto.
\newblock {Cooperative dynamics of oscillator community: a study based on
  lattice of rings}.
\newblock {\em Prog. Theor. Phys. Suppl.}, Suppl. 79:223--240, 1984.

\bibitem{Kuramoto:1991ja}
Y.~Kuramoto.
\newblock {Collective synchronization of pulse-coupled oscillators and
  excitable units}.
\newblock {\em Physica D}, 50(1):15--30, May 1991.

\bibitem{Kuramoto:1997kd}
Y.~Kuramoto.
\newblock {Phase- and center-manifold reductions for large populations of
  coupled oscillators with application to non-locally coupled systems}.
\newblock {\em Int. J. Bifurcat. Chaos}, 7(4):789--805, 1997.

\bibitem{Lohmiller:1998to}
W.~Lohmiller and J.-J.~E. Slotine.
\newblock {On contraction analysis for non-linear systems}.
\newblock {\em Automatica}, 34(6):683--696, June 1998.

\bibitem{Mauroy:2011th}
A.~Mauroy.
\newblock {\em {On the dichotomic collective behaviors of large populations of
  pulse-coupled firing oscillators}}.
\newblock PhD thesis, University of Li{\`e}ge, Li{\`e}ge, Belgium, Oct. 2011.

\bibitem{Mauroy:2010wca}
A.~Mauroy, J.~M. Hendrickx, A.~Megretski, and R.~Sepulchre.
\newblock {Global analysis of firing maps}.
\newblock In {\em Proc. 19th Int. Symp. Mathematical Theory Networks and
  Systems}, pages 1775--1782, Budapest, Hungary, July 2010.

\bibitem{Mauroy:2008dp}
A.~Mauroy and R.~Sepulchre.
\newblock {Clustering behaviors in networks of integrate-and-fire oscillators}.
\newblock {\em Chaos}, 18(3):037122, 2008.

\bibitem{Mauroy:2011ux}
A.~Mauroy and R.~Sepulchre.
\newblock {Global analysis of a continuum model for monotone pulse-coupled
  oscillators}.
\newblock {\em arXiv}, math.AP, Feb. 2011.

\bibitem{Mauroy:2012ts}
A.~Mauroy and R.~Sepulchre.
\newblock {Contraction of monotone phase-coupled oscillators}.
\newblock {\em arXiv}, math.DS, May 2012.

\bibitem{Mirollo:1990ft}
R.~E. Mirollo and S.~H. Strogatz.
\newblock {Synchronization of pulse-coupled biological oscillators}.
\newblock {\em SIAM J. Appl. Math.}, 50(6):1645--1662, 1990.

\bibitem{Moreau:2005km}
L.~Moreau.
\newblock {Stability of multiagent systems with time-dependent communication
  links}.
\newblock {\em IEEE Trans. Autom. Control}, 50(2):169--182, Feb. 2005.

\bibitem{Nagumo:1962iz}
J.~Nagumo, S.~Arimoto, and S.~Yoshizawa.
\newblock {An active pulse transmission line simulating nerve axon}.
\newblock {\em Proc. IRE}, 50(10):2061--2070, 1962.

\bibitem{Nijmeijer:1997kr}
H.~Nijmeijer and I.~Mareels.
\newblock {An observer looks at synchronization}.
\newblock {\em IEEE Trans. Circuits Syst. I}, 44(10):882--890, Oct. 1997.

\bibitem{Nijmeijer:2003uy}
H.~Nijmeijer and A.~Rodr{\'\i}guez-Angeles.
\newblock {\em {Synchronization of Mechanical Systems}}.
\newblock World Scientific, 2003.

\bibitem{Nunez:2012um}
F.~Nunez, Y.~Wang, A.~R. Teel, and F.~J. Doyle~III.
\newblock {Bio-inspired synchronization of non-identical pulse-coupled
  oscillators subject to a global cue and local interactions}.
\newblock In {\em Proc. 4th IFAC Conf. Analysis and Design of Hybrid Systems},
  pages 115--120, June 2012.

\bibitem{Okuda:1993dj}
K.~Okuda.
\newblock {Variety and generality of clustering in globally coupled
  oscillators}.
\newblock {\em Physica D}, 63(3--4):424--436, Mar. 1993.

\bibitem{Olami:1992tv}
Z.~Olami, H.~J.~S. Feder, and K.~Christensen.
\newblock {Self-organized criticality in a continuous, nonconservative cellular
  automaton modeling earthquakes.}
\newblock {\em Phys. Rev. Lett.}, 68(8):1244--1247, Feb. 1992.

\bibitem{Orosz:2009wx}
G.~Orosz, J.~Moehlis, and P.~Ashwin.
\newblock {Designing the dynamics of globally coupled oscillators}.
\newblock {\em Prog. Theor. Phys.}, 122(3):611--630, Sept. 2009.

\bibitem{Ostborn:2002wh}
P.~{\"O}stborn.
\newblock {Phase transition to frequency entrainment in a long chain of
  pulse-coupled oscillators.}
\newblock {\em Phys. Rev. E}, 66(1 Pt 2):016105, July 2002.

\bibitem{Pavlov:2004tb}
A.~V. Pavlov.
\newblock {\em {The output regulation problem: a convergent dynamics
  approach}}.
\newblock PhD thesis, Technische Universiteit Eindhoven, 2004.

\bibitem{Pavlov:2005tu}
A.~V. Pavlov, N.~van~de Wouw, and H.~Nijmeijer.
\newblock {Convergent systems: analysis and synthesis}.
\newblock In {\em Control and Observer Design for Nonlinear Finite and Infinite
  Dimensional Systems}, pages 131--146. Springer-Verlag, Berlin/Heidelberg,
  Germany, 2005.

\bibitem{Peskin:1975wc}
C.~S. Peskin.
\newblock {\em {Mathematical Aspects of Heart Physiology}}.
\newblock Courant Institute of Mathematical Sciences, New York University, New
  York, NY, 1975.

\bibitem{Pikovsky:2001et}
A.~Pikovsky, M.~Rosenblum, and J.~Kurths.
\newblock {\em {Synchronization: A Universal Concept in Nonlinear Sciences}},
  volume~12 of {\em Cambridge Nonlinear Science Series}.
\newblock Cambridge University Press, Cambridge, United Kingdom, 2001.

\bibitem{Rhouma:2001kl}
M.~B.~H. Rhouma and H.~Frigui.
\newblock {Self-organization of pulse-coupled oscillators with application to
  clustering}.
\newblock {\em IEEE Trans. Pattern Anal. Mach. Intell.}, 23(2):180--195, Feb.
  2001.

\bibitem{Russo:2010kx}
G.~Russo, M.~di~Bernardo, and E.~D. Sontag.
\newblock {Stability of networked systems: a multi-scale approach using
  contraction}.
\newblock In {\em Proc. 49th IEEE Conf. Decision and Control}, pages
  6559--6564, Atlanta, GA, Dec. 2010.

\bibitem{Scardovi:2010gx}
L.~Scardovi, M.~Arcak, and E.~D. Sontag.
\newblock {Synchronization of interconnected systems with applications to
  biochemical networks: an input--output approach}.
\newblock {\em IEEE Trans. Autom. Control}, 55(6):1367--1379, June 2010.

\bibitem{Senn:2000tx}
W.~Senn and R.~Urbanczik.
\newblock {Similar nonleaky integrate-and-fire neurons with instantaneous
  couplings always synchronize}.
\newblock {\em SIAM J. Appl. Math.}, 61(4):1143--1155, 2000.

\bibitem{Sepulchre:2006vk}
R.~Sepulchre.
\newblock {Oscillators as systems and synchrony as a design principle}.
\newblock In {\em Current Trends in Nonlinear Systems and Control: In Honor of
  Petar Kokotovi{\'c} and Turi Nicosia}. Birkh{\"a}user, Boston, MA, 2006.

\bibitem{Slotine:2005vv}
J.-J.~E. Slotine and W.~Wang.
\newblock {A study of synchronization and group cooperation using partial
  contraction theory}.
\newblock In {\em Cooperative Control}, pages 207--228. Springer-Verlag,
  Berlin/Heidelberg, Germany, 2005.

\bibitem{Sontag:2010js}
E.~D. Sontag.
\newblock {Contractive systems with inputs}.
\newblock In {\em Perspectives in Mathematical System Theory, Control, and
  Signal Processing}, pages 217--228. Springer, Berlin/Heidelberg, Germany,
  2010.

\bibitem{Sontag:2008tr}
E.~D. Sontag and M.~Arcak.
\newblock {Passivity-based stability of interconnection structures}.
\newblock In {\em Recent Advances in Learning and Control}, pages 195--204.
  Springer-Verlag, London, England, 2008.

\bibitem{Stan:2007jy}
G.-B. Stan and R.~Sepulchre.
\newblock {Analysis of interconnected oscillators by dissipativity theory}.
\newblock {\em IEEE Trans. Autom. Control}, 52(2):256--270, Feb. 2007.

\bibitem{Strogatz:2000wx}
S.~H. Strogatz.
\newblock {From Kuramoto to Crawford: exploring the onset of synchronization in
  populations of coupled oscillators}.
\newblock {\em Physica D}, 143(1-4):1--20, Sept. 2000.

\bibitem{Strogatz:2003tm}
S.~H. Strogatz.
\newblock {\em {Sync: the Emerging Science of Spontaneous Order}}.
\newblock Hyperion, New York, NY, 2003.

\bibitem{Timme:2002vu}
M.~Timme, F.~Wolf, and T.~Geisel.
\newblock {Coexistence of regular and irregular dynamics in complex networks of
  pulse-coupled oscillators.}
\newblock {\em Phys. Rev. Lett.}, 89(25):258701, Dec. 2002.

\bibitem{Timme:2002uk}
M.~Timme, F.~Wolf, and T.~Geisel.
\newblock {Prevalence of unstable attractors in networks of pulse-coupled
  oscillators.}
\newblock {\em Phys. Rev. Lett.}, 89(15):154105, Oct. 2002.

\bibitem{Timme:2004tx}
M.~Timme, F.~Wolf, and T.~Geisel.
\newblock {Topological speed limits to network synchronization}.
\newblock {\em Phys. Rev. Lett.}, 92(7):074101, Feb. 2004.

\bibitem{VanderPol:1920um}
B.~van~der Pol.
\newblock {A theory of the amplitude of free and forced triode vibrations}.
\newblock {\em Radio Rev.}, 1:701--710, 1920.

\bibitem{vanVreeswijk:1996wb}
C.~van Vreeswijk.
\newblock {Partial synchronization in populations of pulse-coupled
  oscillators}.
\newblock {\em Phys. Rev. E}, 54(5):5522--5537, Nov. 1996.

\bibitem{vanVreeswijk:1994wy}
C.~van Vreeswijk, L.~F. Abbott, and G.~B. Ermentrout.
\newblock {When inhibition not excitation synchronizes neural firing}.
\newblock {\em J. Comput. Neurosci.}, 1(4):313--321, 1994.

\bibitem{Wang:2004jm}
W.~Wang and J.-J.~E. Slotine.
\newblock {On partial contraction analysis for coupled nonlinear oscillators}.
\newblock {\em Biol. Cybern.}, 92(1):38--53, Dec. 2004.

\bibitem{Watanabe:1994hh}
S.~Watanabe and S.~H. Strogatz.
\newblock {Constants of motion for superconducting Josephson arrays}.
\newblock {\em Physica D}, 74(3-4):197--253, July 1994.

\bibitem{Winfree:1967vf}
A.~T. Winfree.
\newblock {Biological rhythms and the behavior of populations of coupled
  oscillators}.
\newblock {\em J. Theor. Biol.}, 16(1):15--42, July 1967.

\bibitem{Winfree:1980ue}
A.~T. Winfree.
\newblock {\em {The Geometry of Biological Time}}, volume~8 of {\em
  Biomathematics}.
\newblock Springer-Verlag, New York, NY, 1st edition, 1980.

\bibitem{YouTube:2011uh}
{YouTube}.
\newblock {Sync of metronomes} [online].
\newblock Mar. 2011.
\newblock URL: \url{http://www.youtube.com/watch?v=gFnVmuU8\textunderscore Lg}.

\bibitem{Zillmer:2007kl}
R.~Zillmer, R.~Livi, A.~Politi, and A.~Torcini.
\newblock {Stability of the splay state in pulse-coupled networks}.
\newblock {\em Phys. Rev. E}, 76(4 Pt 2):046102, Oct. 2007.

\end{thebibliography}
%\bibliography{./bib/am-bibfile,./bib/ps-bibfile,./bib/rs-bibfile}

\end{document}